# Non-LTE models for synthetic spectra of type Ia supernovae

## III. An accelerated lambda iteration procedure for the mutual interaction of strong spectral lines in SN Ia models with and without energy deposition

A. W. A. Pauldrach, T. L. Hoffmann, and P. J. N. Hultzsch

Universitäts-Sternwarte München, Scheinerstr. 1, 81679 München, Germany
e-mail: uh10107@usm.lmu.de, hoffmann@usm.lmu.de, pjnh@usm.lmu.de

### ABSTRACT

*Context.* In type Ia supernova (SN Ia) envelopes a huge number of lines of different elements overlap within their thermal Doppler widths, and this problem is exacerbated by the circumstance that up to 20% of these lines can have a line optical depth greater than 1. The stagnation of the lambda iteration in such optically thick cases is one of the fundamental physical problems *inherent* in the iterative solution of the non-LTE problem, and the failure of a lambda iteration to converge is a point of crucial importance whose physical significance must be understood completely.

*Aims.* We discuss a general problem related to radiative transfer under the physical conditions of supernova ejecta that involves a failure of the usual non-LTE iteration scheme to converge when multiple strong opacities belonging to different physical transitions come together, similar to the well-known situation where convergence is impaired even when only a single process attains large optical depths. The convergence problem is independent of the chosen frequency and depth grid spacing, independent of whether the radiative transfer is solved in the comoving or observer's frame, and independent of whether a common complete-linearization scheme or a conventional accelerated lambda iteration (ALI) is used. The problem appears when all millions of line transitions required for a realistic description of SN Ia envelopes are treated in the frame of a comprehensive non-LTE model. The only way out of this problem is a complete-linearization approach which considers all ions of all elements simultaneously, or an adequate generalization of the established ALI technique which accounts for the mutual interaction of the strong spectral lines of different elements and which thereby unfreezes the "stuck" state of the iteration.

*Methods.* The physics of the atmospheres of SN Ia are strongly affected by the high-velocity expansion of the ejecta, dominating the formation of the spectra at all wavelength ranges. Thus, hydrodynamic explosion models and realistic model atmospheres that take into account the strong deviation from local thermodynamic equilibrium are necessary for the synthesis and analysis of the spectra. In this regard one of the biggest challenges we have found in the modeling of the radiative transfer in SN Ia is the fact that the radiative energy in the UV has to be transferred only via spectral lines into the optical regime in order to be able to leave the ejecta. However, convergence of the model toward a state where this is possible is impaired when using the standard procedures. We report on improvements in our approach of computing synthetic spectra for SN Ia with respect to (i) an improved and sophisticated treatment of many thousands of strong lines that interact intricately with the "pseudo-continuum" formed entirely by Doppler-shifted spectral lines, (ii) an improved and expanded atomic database, and (iii) the inclusion of energy deposition in the ejecta arising from the radioactive decay of mostly $^{56}$Ni and $^{56}$Co.

*Results.* We show that an ALI procedure we have developed for the mutual interaction of strong spectral lines appearing in the atmospheres of SNe Ia solves the longstanding problem of transferring the radiative energy from the UV into the optical regime. Our new method thus constitutes a foundation for more refined models, such as those including energy deposition. In this regard we further show synthetic spectra obtained with various methods adopted for the released energy and compare them to observations. In detail we discuss applications of the diagnostic technique by example of a standard type Ia supernova, where the comparison of calculated and observed spectra revealed that in the early phases the consideration of the energy deposition within the spectrum-forming regions of the ejecta does not qualitatively alter the shape of the emergent spectra.

*Conclusions.* The results of our investigation lead to an improved understanding of how the shape of the spectrum changes radically as function of depth in the ejecta, and show how different emergent spectra are formed as a result of the particular physical properties of SNe Ia ejecta and the resulting peculiarities in the radiative transfer. This knowledge provides an important insight into the process of extracting information from observed SN Ia spectra, since these spectra are a complex product of numerous unobservable SN Ia spectral features which are thus analyzed in parallel to the observable SN Ia spectral features.

**Key words.** radiative transfer – supernovae: general – supernovae: individual (SN 1992A)

## 1. Introduction

Realistic models and synthetic spectra of good quality are required for supernova of type Ia in order to tackle the question whether these objects are "standard candles" in a cosmological sense[1]. Because of the complex physical processes involved in the explosion mechanism, the nucleosynthesis, and the radiation transport the development of such models is however not a simple task.

Type Ia supernovae have become an important tool for the determination of the cosmological parameters (Riess et al. 1998; Perlmutter et al. 1999; Riess et al. 2001; Tonry et al. 2003), since their exceptional brightness makes them observable even at large cosmological distances. Using brightness measurements for distance determination requires knowledge of the absolute luminosities of the objects, and the application of SN Ia for this task relies on the crucial assumption that the objects observed at high redshifts have the same properties as those in the nearby universe. Thus, for the observed brightness differences of distant SN Ia the same calibration methods are applied as for the local objects. To develop a realistic model which describes the explosion mechanism and reproduces the light curves and spectra of SN Ia in detail, including the observed intrinsic variability, is therefore of great importance.

The conceptual basis of such a model is a carbon-oxygen white dwarf (WD) which reaches a mass close to the Chandrasekhar mass ($M_{\rm Ch} \approx 1.4 M_\odot$). As a consequence, carbon burning ignites close to the center of the star due to compressional heating, and after a period of a few thousand years of quiet burning a thermonuclear runaway disrupts the star (Iben & Tutukov 1984; Webbink 1984; Woosley et al. 1984; Han & Podsiadlowski 2004). Although this is the currently favored mechanism for a thermonuclear SN Ia explosion, the progenitor scenario has not been clarified yet. In the more or less accepted "single degenerate" scenario the WD accretes mass from a red-giant companion star, but the potential progenitor systems that have been found in recent years clearly indicate that their numbers are too low to explain the observed SN Ia rates (Cappellaro et al. 1999; Maoz 2008; Gilfanov & Bogdán 2010). On the other hand, Pauldrach (2005) has pointed out, based on the results of Pauldrach et al. (2004),[2] that SN Ia progenitors are very likely connected to a subgroup of central stars of planetary nebulae (CSPNs) (see also Kaschinski et al. 2012 and Kaschinski et al. 2013).

But the progenitor scenario is not the only subject of a lively debate. Another one is the explosion process itself. As a general mechanism both deflagration and detonation scenarios are discussed. In the first case a subsonic "flame" (deflagration wave) is ignited near the center of the star, which travels outward, burning part of the star to nuclear statistical equilibrium, and the star expands. (Since the flame propagates subsonically during this burning phase the star does not explode.) This process allows partial burning of C and O to intermediate mass elements (Si, S, Mg, Ca) which thus dominate the composition, but the observed iron-group elements (Fe, Co, Ni) are not sufficiently produced. In contrast, a prompt, supersonic detonation of the star generates primarily iron-group elements, but this is also in contradiction to the observed composition. Thus, the "explosion" cannot continue subsonically for ever (Nomoto et al. 1984; Woosley

---

et al. 1984; Niemeyer & Hillebrandt 1995; Reinecke et al. 2002; Röpke & Hillebrandt 2005; Röpke 2005) and a (yet unknown) mechanism has to trigger the transition of the subsonic deflagration into a supersonic detonation ("delayed detonation transition") (Hoeflich & Khokhlov 1996; Iwamoto et al. 1999). Although the resulting composition of this model is generally not in contradiction to observations, this model has a somewhat artificial character, since the occurrence of the delayed detonation is based on an ad-hoc assumption. As realistic radiative transfer models provide the link between explosion models and the observations, they perforce assume the position of arbiter between the different explosion scenarios, and their application will make it possible to judge which simulations describe the explosions correctly.

But the radiative transfer must be modelled in some detail in order to establish observational constraints for the explosion models. To simulate and understand the processes involved in SN Ia envelopes different numerical approaches have been developed to describe the radiative transfer and the evolution of the light curves: based on a variety of approaches and involving different levels of complexity models have been developed by several groups (Branch et al. 1985; Mazzali et al. 1993; Mazzali & Lucy 1993; Eastman & Pinto 1993; Höflich et al. 1995; Nugent et al. 1995a; Pauldrach et al. 1996; Nugent et al. 1997; Lentz et al. 2001; Höflich 2005; Stehle et al. 2005; Sauer et al. 2006; Baron et al. 2006; Kasen et al. 2006). However, with a view to the purpose of the specific codes various simplifications have been applied in the past, as not all approaches were intended to provide a comprehensive description of the time-dependent spectra, including the detailed statistical equilibrium of all relevant elements. Highly parameterized models, for example, which implement a simplified treatment of physical processes to achieve short run-times seem to be suitable for the comparative analysis of a large number of observed spectra, while more realistic (but computationally more expensive) models are required for a deeper understanding of the physical effects leading to specific observed properties.

Our focus lies in a sophisticated description of the SN Ia spectra with respect to high spectral resolution (Pauldrach et al. 1996; Sauer et al. 2006), in order to quantify the observable physical properties of SN Ia accurately, in particular also with the intent of determining possible observational features that might be used to discriminate between different hydrodynamic explosion models or at least particular aspects of these scenarios. (Judging the validity of hydrodynamic explosion models can only be performed by using radiative transfer models that include a very detailed treatment of the relevant physical processes; such detailed models may also be used to validate or invalidate specific simplifying assumptions used in less elaborate models.) In this work we present a comprehensively improved method, which is based on radioactive energy rates obtained from the decay chain $^{56}$Ni $\rightarrow$ $^{56}$Co $\rightarrow$ $^{56}$Fe and deposited in the ejecta, and used to obtain a consistent solution of the full non-LTE detailed statistical equilibrium along with a detailed solution of the radiative transfer in the observer's frame[3]. As the physics of the atmospheres of SN Ia are strongly affected by the velocity fields in the expanding ejecta, dominating the spectra at all wavelength ranges, a decisive improvement of our approach to compute synthetic spectra for SN Ia also regards a sophisti-

---

cated treatment of the thousands of strong lines which interact in an interwoven way with the "pseudo-continuum" that is entirely formed by Doppler-shifted spectral lines.

In the following we will first introduce the theoretical basis of our simulations, describing the general concept used to model the physical state of the gas and the transport of radiation (Sect. 2). In Sect. 3 we discuss our numerical approach to treating the time-dependence of the radioactive decay and the transport and deposition of the released energy. In Sect. 4 we discuss important details of our approach to calculate synthetic spectra for SN Ia and present tests comparing our results to observations. In Sect. 5 we explore the question of whether the spectra of our current models of SN Ia are already realistic enough to be used for diagnostic purposes, and we interpret and summarize our results along with an outlook in Sect. 6.

## 2. Overview of the method

To be able to investigate the systematic errors in the cosmological quantities derived from SN Ia observations, it is essential first to develop a detailed understanding of the physics involved in the explosion process. This in turn requires a detailed understanding of the production and the transport of radiation in the ejecta, in order to link the observable quantities – the light curves and spectra of SN Ia – to the details of the explosions.

To set up the radiative transfer models correctly basic information deduced from observations is required, and the observations show that the evolution of a type Ia supernova can be divided into different phases.

At early epochs – before and shortly after maximum of the light curve – SN Ia exhibit a spectrum which is dominated by a few very broad absorption features embedded in a non-thermal continuum. While the absorption features are mostly the product of blends of several lines, the "continuum" is formed by the overlap of a large number of Doppler-broadened metal lines producing a so called "pseudo-continuum" – this defines the photospheric phase. The true continuum opacities and emissivities which determine the overall shape of stellar spectra are only of minor significance in supernovae. In the photospheric phase the ejecta can nevertheless be treated analogously to hot stars with expanding, extended atmospheres, and similar concepts for the solution of the radiative transfer can be applied. Because of the increasing optical depths in deeper layers of the "pseudo-continuum", the radiative transfer needs to be considered only in the photosphere and in the thin envelope above this layer.

At later epochs, the "pseudo-continuum photosphere" recedes deeper and deeper into the ejecta and finally disappears when the ejecta become transparent – this defines the nebular phase. In these late phases the concepts developed for modelling gaseous nebulae can basically be adopted.

The most important difference between SN Ia and stars in the context of atmospheric modelling concerns the production of the radiation. In contrast to the atmospheres of stars, in SN Ia the radiation is generated within the expanding layers themselves, by the deposition of energy resulting from the (time-dependent) decay of $^{56}Ni \rightarrow ^{56}Co$ and $^{56}Co \rightarrow ^{56}Fe$ (Colgate & McKee 1969; see also Figs. 2 and 3). This behavior of course modifies the transport of radiation, and a radiative transfer model for SN Ia which includes the treatment of energy deposition by the decay products of $^{56}Ni$ and $^{56}Co$ requires in principle a completely time-dependent consistent solution of the populations of all atomic levels, the radiative continuum and line transfer, and the released energy. However, as in the photospheric phase the radiative transfer needs to be considered only for the envelope where the photon escape timescale is generally much shorter than the expansion timescale, a steady-state solution is sufficient. With this in mind, we will specify further our concept for calculating consistent atmospheric models of SN Ia in Sect. 2.2.

But before we discuss in detail the current status of our treatment of expanding atmospheres of supernovae, we will first summarize in the next two sub-sections the general concept of our procedure.

### 2.1. The general principle of consistent SN Ia models

In order to determine observable physical properties of supernovae (such as abundance distributions, velocity and density structures, ionization and excitation states of the ions, and energy deposition rates) via quantitative spectroscopy one has to calculate snapshots of the time-dependent synthetic spectra of these objects. The calculation of these spectra requires as a basis results from the treatment of the physics of the evolution of the explosion mechanism which are expressed by the luminosity, the epoch, the hydrodynamical structure, and the chemical composition.

**Luminosity.** The luminosity of SN Ia results primarily from the energy that is deposited by $\gamma$-photons originating from the decay of radionuclides synthesized in the explosion, and the photons emitted at a given time consist not just of the photons created at that particular instant (via the processes following nuclear decay), but also of the photons which have been generated at earlier times and trapped by the large opacities $\chi_\nu$ (cf. Arnett 1982; Hoeflich & Khokhlov 1996; Nugent et al. 1997; Pinto & Eastman 2000) – the photons can not immediately penetrate through the envelope because of their small mean free path lengths $l_\nu = \chi_\nu^{-1}$; thus, the release of the locally trapped radiative energy requires an increase of $l_\nu$ which takes place along with the expansion of the object.

Since the luminosity is fed by the energy deposition, it is largely dependent on the radial distribution of the respective elements in the ejecta (see below). This behavior of course has implications for the method applied to simulate the envelope, since it is not the total luminosity which originates from below the boundary of the simulation volume, as is the case for stellar atmospheres. Instead, the incoming luminosity at the inner boundary must be specified to account for the radiative energy deposited below the simulation volume, and the energy deposition within the simulation volume must be treated explicitly. If this physical behavior is described correctly, the time-dependent luminosity can be derived from the mass of $^{56}Ni$ predicted for the specific SN Ia explosion (see below).

**Epoch.** As the SN Ia ejecta are in a state of homologous expansion, the epoch $t$, the time after explosion, is an important model parameter not only because the nuclear decay rates feeding the luminosity are explicitly time-dependent, but also because the density–radius relation changes with time (see below). Due to the trapping of photons described above and the cooling of the gas by the expansion, the emission of light is delayed, and SN Ia are thus not immediately visible directly after explosion. The epoch is therefore not a directly measurable quantity, but can be inferred with the help of light-curve simulations.

**Hydrodynamical structure.** The relationship between radius, density, and velocity at a specific time after explosion (i.e., a given epoch) is presented by the hydrodynamical structure. This structure is obtained from the expansion of the envelope, which is ballistic, and therefore a homologous structure of the ejecta is a safe assumption. The radius $r$ and velocity $v$ are thus related via $r = vt$, and a given density structure $\rho_0(r, t_0)$ at epoch $t_0$ scales with $t$ as $\rho(r, t) = \rho_0((t_0/t)r, t_0)(t_0/t)^3$. (In order to describe the radial structure of the ejecta the velocity is usually used as the basic coordinate, since it is independent of the epoch.) The natural assumption to fix the hydrodynamical quantities $r$, $v$, and $\rho$ for a snapshot calculation of the non-LTE model[4] then also leads to a relationship of the optical depth scale $\tau$ – which defines the important quantity of the "photospheric radius", for instance – and the density.

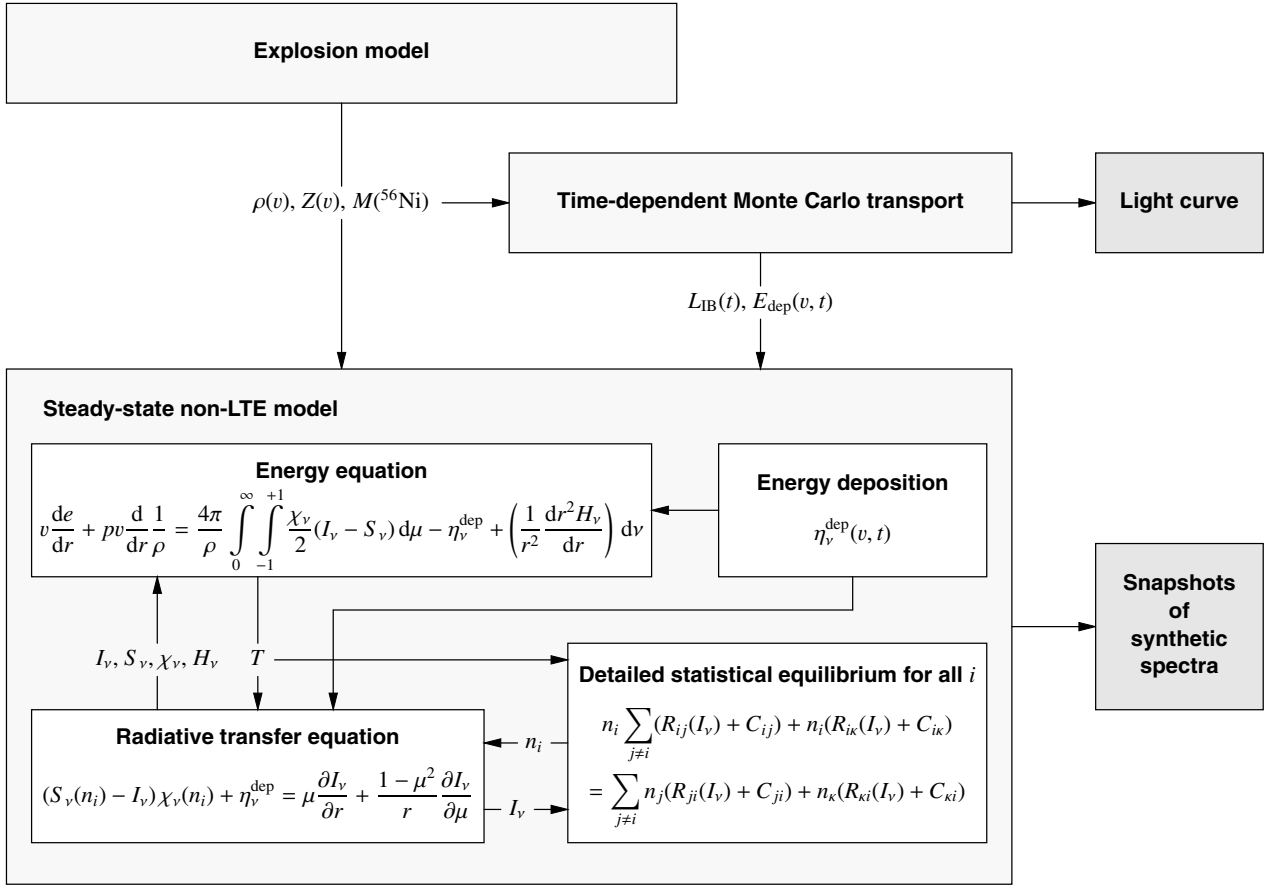

**Fig. 1.** Sketch of our procedure to calculate synthetic spectra and light curves from the results of SN Ia explosion model calculations ($M(^{56}\text{Ni})$ is the total mass of $^{56}$Ni and $Z$ represents the chemical composition of the supernova ejecta which in general has a radial stratification – cf. Sect. 2). *In a first step the time-dependent radiative transfer is solved*, determining the energy deposition rates $E_{\text{dep}}(v, t)$ and the luminosity $L_{\text{IB}}(t)$ at the inner boundary of the non-LTE model along with the light curve. ($\eta_\nu^{\text{dep}}(v, t)$ is the emission coefficient resulting from the energy deposition rates (cf. Sect. 3). $L_{\text{IB}}(t)$, the luminosity which is used for the inner boundary condition of the non-LTE model (cf. Sect. 4), is obtained from the solution of the time dependent radiative transfer of the deeper layers.) *For the outer part of the ejecta a snapshot calculation for a stationary non-LTE model atmosphere is performed* using the energy deposition rates and the luminosity from the time-dependent radiative transfer. As a result this simulation yields the temperature structure, the occupation numbers along with the ionization structure, the radiation field, and the synthetic spectrum. The "non-LTE model"-part shows the non-linear system of integro-differential equations that form the basis of stationary atmospheric models, where the detailed statistical equilibrium determines via the collisional ($C_{ij}$) and radiative ($R_{ij}$) transition rates the occupation numbers $n_i$ of the atomic levels $i$, $j$; the spherical radiative transfer equation yields the radiation field ($\chi_\nu$ and $S_\nu$ are the total opacity and source function of all microphysical processes that are explicitly considered) expressed by the specific intensity $I_\nu$, the mean intensity $J_\nu$ and the Eddington flux $H_\nu$; and the microscopic energy equation of matter[5] gives the temperature structure $T(r)$ within the simulated part of the supernova ejecta (cf. Sect. 4). $r$ is the radial coordinate, $\rho$ the mass density, $v$ the velocity, $p$ the gas pressure, and $e$ the internal energy; the left hand side of the energy equation contains the change of the internal energy and the adiabatic cooling term and the right hand side is the negative of the 0th moment of the transfer equation describing the total radiation field.

**Chemical composition.** The chemical composition in supernova ejecta is in general a function of the radius. To specify the structures of the corresponding abundances the composition is either taken from the nucleosynthesis calculations of an explosion model, or it is determined via a fit of the synthetic to the observed spectrum. Thus, from a comparison of the calculated and fitted composition of a SN Ia the predictions of specific explosion models can be tested in principle.

### 2.2. The concept for consistent atmospheric models of SN Ia

It is primarily the high radiation energy density acting on the environment of low matter density surrounding the SN Ia center of explosion, as well as the energy input by γ-rays and positrons within this envelope, which is responsible for the circumstance that the simplifying assumptions of LTE are not applicable to SN Ia. Moreover, supernovae are time-dependent not only with respect to the explosion scenario that leads to non-stationary expansion with very high velocities (up to 25 000 km/s), but also with regard to the energy released into the ejecta by the profoundly time-dependent radioactive decay of $^{56}$Ni and $^{56}$Co (cf. Fig. 2). To account for this behavior we have in our approach split the description of the time-dependent deposition of radiative energy based on γ photons, which originate from the $^{56}$Ni and $^{56}$Co decay and which are initially trapped in the highly opaque expanding material, finally leading after being released to the characteristic shapes of the light curves (cf. Fig. 3), from the description of the formation of the spectra (cf. Fig. 1), which originate only from the outer parts of the ejecta where the material is less opaque and the radiation can emerge (as in these parts of the atmospheres the time scales for interaction of photons with matter are much smaller than the expansion time scales, steady state models can be assumed for the calculation of the spectra).

---

[4] The term non-LTE refers to detailed modelling of the statistical equilibrium of the occupation numbers of the atomic levels, obtained from actually solving the system of rate equations, without assuming the approximation of local thermodynamic equilibrium (LTE) in which the level populations would follow a Saha-Boltzmann distribution at the local temperature and density.

[5] The microscopic energy equation is usually obtained by subtracting the dot product of the Euler equation with $\boldsymbol{v}$ from the general energy equation of matter.

**Time-dependent energy deposition.** The observed luminosity evolution of SN Ia (Fig. 3) is very slow compared to the (hypothetical) case that the luminosity is powered only by the thermal energy from the explosion. This is recognized especially at epochs later than 100 days where the light curves decline almost linearly with a rate of ∼ 1 mag per 100 days, pointing to a delayed energy input to the outer layers of the envelopes and a subsequently further delayed release of radiative energy from these layers. That the radioactive decay of the unstable elements $^{56}$Ni and $^{56}$Co is the primary source of energy can moreover be deduced from the – on a logarithmic scale – linear tail of the light curves, because these processes favor an exponential law for the time-dependent release of energy. To account for this behavior focus will be given to a correct description of the time dependent energy production rates, their spatial and frequential distribution within the atmosphere, and the transport of the photons resulting from the energy production rates and their interaction with matter. A detailed description of the procedure for calculating the time-dependent energy deposition rates is given in Sect. 3. These deposition rates are required for the calculation of realistic synthetic spectra, and their accuracy is verified by a comparison of the computed synthetic light curve with corresponding observed ones.

**Non-LTE models and synthetic spectra.** The energy deposition rates calculated at specific epochs comprise one of the factors required for the non-LTE radiation transport and spectral synthesis models which treat the outer parts of the ejecta as "snapshots" in time. These calculations we perform for times around maximum and later, and only for the envelope which starts to become optically thin in at least some spectral regions. To calculate synthetic spectra (on basis of stationary and spherically symmetric envelopes with "photospheres" as lower boundaries), detailed physics of different nature must be implemented in the required procedure. This has been done for the code used, which originally had been developed for analyzing spectra of expanding atmospheres of hot stars (Pauldrach 1987; Pauldrach et al. 1990; Pauldrach et al. 1994; Pauldrach et al. 2001; Pauldrach et al. 2012) and which has been adapted to the physical conditions of SN Ia (Pauldrach et al. 1996; Sauer et al. 2006, and Hoffmann et al. 2013) and gaseous nebulae (Hoffmann et al. 2012 and Weber et al. 2013). It provides a consistent solution of the non-LTE rate equations for all relevant elements, and the detailed radiative transfer takes into account all significant sources of opacity and emission on the basis of a proper treatment of line blocking[6] and blanketing[7] effects. The radiation field at the outermost region of the converged model represents the synthetic spectrum of the object, which can in principle be directly compared to observations.

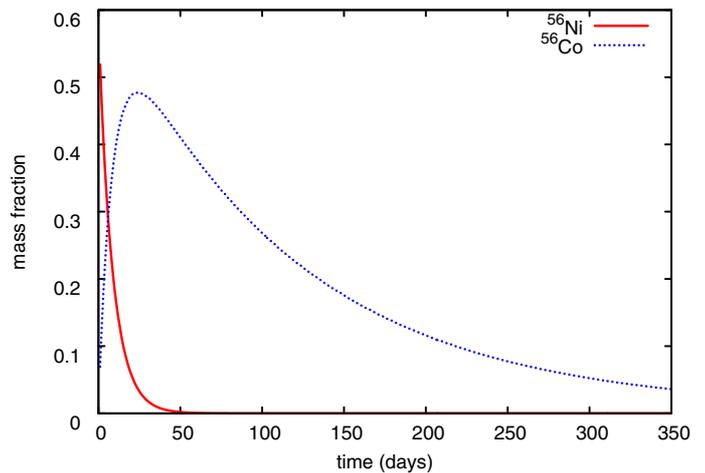

**Fig. 2.** Time-dependent evolution of the mass fraction of $^{56}$Ni and $^{56}$Co. The decay of these nuclei is the dominant source of energy for the light emitted by SN Ia. As the half-life of $^{56}$Ni to $^{56}$Co is very short (6.1 days) this decay is only important in the first 50 days. $^{56}$Co on the other hand decays to the stable nucleus $^{56}$Fe within a much longer half-life (77.3 days) and is therefore the dominant source of energy for a period of 2 to 3 years.

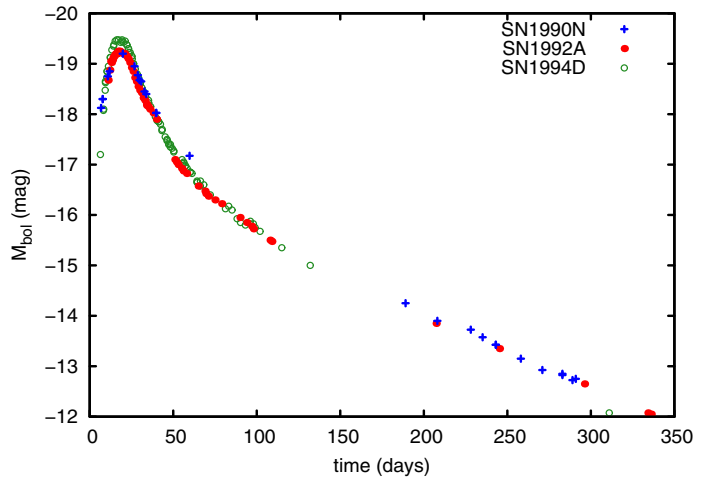

**Fig. 3.** Observed absolute V light curves of three SNe Ia (SN1990N, SN1994D, and SN1992A – Kirshner et al. 1993). Characteristic features of the curves are the steep increase before maximum light (at about 18 days after explosion), the steep decrease until about 50 days after explosion, and the slower decrease later on.

---

[6] The effect of line blocking refers to an attenuation of the radiative flux in the EUV and UV spectral ranges due to the combined opacity of a huge number of metal lines present in SN Ia in these frequency ranges. It drastically influences the ionization and excitation of the radiation field through radiative absorption and scattering processes.

[7] As a consequence of line blocking, only a small fraction of the radiation is re-emitted and scattered in the outward direction, whereas most of the energy is radiated back to deeper layers of the SN Ia, leading there to an increase of the temperature ("backwarming"). Due to this increase of the temperature, more radiation is emitted at lower energies, an effect known as line blanketing.

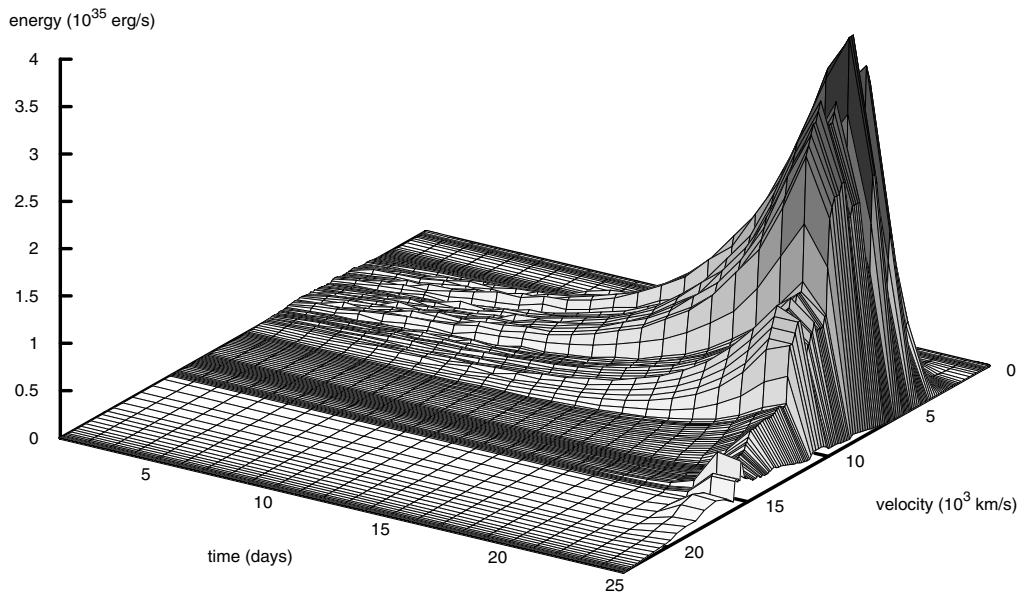

energy ($10^{35}$ erg/s)

velocity ($10^3$ km/s)

time (days)

**Fig. 4.** Spatial and temporal origin of optical photons escaping the supernova ejecta at 25 days after explosion. Most of the energy comes from photons that were created only 2 days before they escape after a comparatively low number of scatterings. While only few photons escape on the same day they have been created, some of the escaping photons were created directly after the explosion and traveled for 25 days through the deeper layers of the atmosphere, suffering many scatterings.

Realistic models for the radiative transfer of SN Ia envelopes represent the crucial link between theoretical explosion models and observations and are therefore an important tool, but the complex physical conditions within the expanding SN Ia ejecta make it extremely difficult to implement models that allow reliable, quantitative analysis. This is the reason why such models, which require at minimum the solution of the full non-LTE problem, are still missing. In Sect. 4 we will describe our approach to construct a more consistent theoretical description of the SN Ia radiation transport.

## 3. Time-dependent radiative transfer

As the brightness of SN Ia originates primarily from the energy deposited by the γ-rays and high-energy positrons generated by the decay of the main products of the nuclear synthesis, $^{56}$Ni and $^{56}$Co (Truran et al. 1967; Colgate & McKee 1969), the production and the transport of this energy in the atmosphere has to be considered explicitly. In principle this means that all relevant microphysical processes influencing the energy range from MeV down to the radio band have to be taken into account. But such a complete time-dependent non-LTE description of the expanding atmosphere is still difficult to realize with present day computers. We have therefore at the present stage decided to focus on a correct treatment of this issue only for the most relevant aspects and to use suitable approximations for those processes which either do not strongly depend on time, or are not strongly affected by detailed non-LTE effects. A high spectral resolution and a detailed non-LTE treatment of the atomic states and the radiative transfer is obviously critical for the computation of the synthetic spectra, since these are formed in the outer layers of the atmospheres where the mean free path lengths of the photons can become large. On the other hand, the origin of the changing luminosities during the early phases of supernovae, caused by the time-dependent nuclear decay chains and the expansions of the ejecta, is to be found deep in the atmospheres where the mean free path lengths of the photons are generally small. In order to simulate the behavior of the luminosities correctly an ex-plicit description of the time-dependence is obviously decisive, but simplifying assumptions with respect to the physical state of the matter – such as the use of mean opacities – are acceptable here because of the small mean free photon path lengths to be found in the layers that need to be considered.

To quantify the amount and distribution of the energy deposited in the envelopes of SNe Ia we use a modified version of the Monte Carlo code described by Cappellaro et al. (1997), which simulates the propagation and absorption of the γ-photons and positrons within the ejecta. To ensure that the approximations used in this calculation are not unrealistic, its predictions are tested via a comparison of the also resulting synthetic light curves with corresponding observed ones.

Our procedure starts from the results of an explosion model which gives the abundance and density stratification as well as the total mass $M(^{56}Ni)$ of nickel. From this we calculate in a first step the amount of energy released in the form of γ-rays and positrons through the nuclear decay chains. In a second step we compute the transport of the γ-photons and positrons until their energy is completely absorbed by the atmosphere. This gives the local energy deposition rates caused by the radioactive decay. As an approximation the complex, multiple-scattering processes of γ-photons dominated by Compton scattering, where low-angle forward scattering with little energy transfer is the most probable (this means that a γ-photon follows a mostly straight line until it undergoes the first large-angle scattering where it looses most of its energy), are described by wavelength-independent opacities.[8] For the positrons we also assume energy-independent interaction cross sections, but we do not assume instant deposition of the positron energy, since late-time light curve fits suggest that the atmosphere eventually becomes transparent even to positrons (Chan & Lingenfelter 1993; Colgate et al. 1997). We thus follow the positrons up to the point they loose their kinetic energy and annihilate. The two γ-photons resulting from this process are then treated as described above.

---

[8]  It has been shown that in a cool supernova atmosphere the scattering interactions are independent of the temperature and ionization state of the gas and therefore the gray transfer provides a very good approximation for the γ-photon deposition (Swartz et al. 1995).

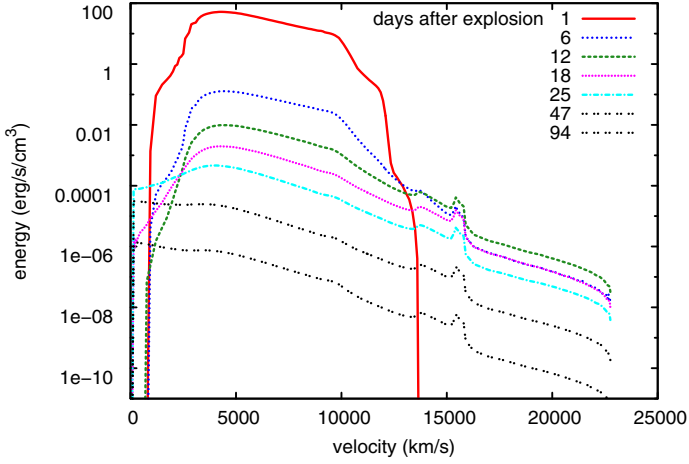

**Fig. 5.** Total deposited energy of γ-photons and positrons as function of depth at different times. In the early phase the energy is deposited locally, since the mean free paths of γ-photons and positrons are very small. 12 days after explosion the γ-photons already reach the outer regions, but not the optically dense center. Only after 25 days does the center also become optically thin for the γ-photons.

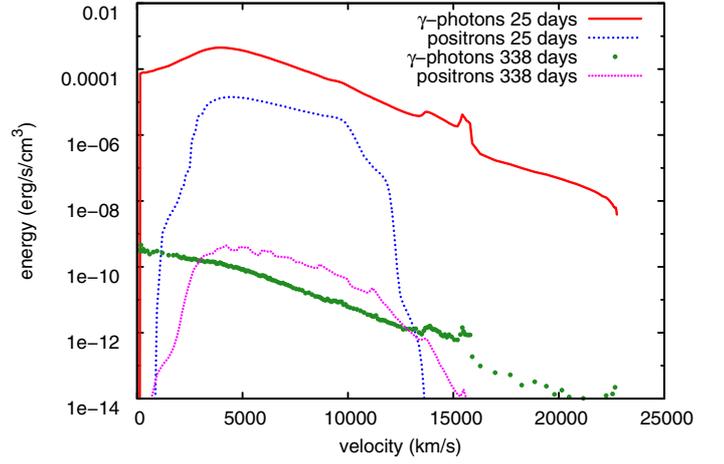

**Fig. 6.** Deposited energy of γ-photons and positrons 25 days and 338 days after explosion. While in the early phase only the deposition of γ-photons is relevant for the brightness of the supernovae, the deposition from the positrons becomes the dominant source of energy in the late phases.

In a third step, the deposited energy from γ-photons and positrons is calculated and converted to optical photons, whose transport through the atmosphere is followed until they escape and contribute to the time-dependent luminosity of the SN Ia. An example of the results of this procedure is given in Fig. 4, which shows the origin of the optical photons that escape the atmosphere 25 days after the explosion. Although most of the escaping photons come from energy deposited 23 days after explosion, a remarkable conclusion from this figure is that a part of these photons have had long travel times, being created already shortly after the explosion. Long travel times of the photons through the ejecta are also the reason why the SN Ia show the highest luminosities 18 days after the explosion (not till then is the atmosphere thin enough for the optical photons to escape), although $^{56}$Ni has a half-life of only 6.1 days. Figure 4 thus gives an example of the time scales to be considered for the random walks of the photons through the SN Ia atmosphere.

The amount of energy released through the radioactive decay chains in the transition of $^{56}$Ni to $^{56}$Co is 1.71 MeV (γ-photon; Burrows & The 1990), and 3.67 MeV in the transition of $^{56}$Co to $^{56}$Fe. In the latter case most of the decay energy is also released in the form of a γ-photon, but a considerable amount (19%) is released as positrons, with about 3.5% of the total energy from the $^{56}$Co decay being kinetic energy of the positron (Arnett 1979; Axelrod 1980). The corresponding energy production coefficients are listed in Table 1.

**Table 1.** Half-lifes $\tau$ and energy production coefficients $e_p$ for the radioactive nuclei considered for the energy deposition and light curve calculations.

| Element | $\tau$ | $e_p$ |
|---|---|---|
| | (days) | ($10^9$ erg/g/s) |
| $^{56}$Ni | 6.08 | 48.6 |
| $^{57}$Ni | 1.48 | 299.0 |
| $^{56}$Co | 77.27 | 8.2 |
| $^{57}$Co | 271.79 | 0.4 |

With the values shown in Table 1 the total energy production rate

$$E_p(t) = E_p(^{56}\text{Ni}, t) + E_p(^{57}\text{Ni}, t) + E_p(^{56}\text{Co}, t) + E_p(^{57}\text{Co}, t)$$

can be calculated. Here $E_p(^{56,57}\text{Ni}, t)$ are the energy production rates from decaying $^{56}$Ni and $^{57}$Ni

$$E_p(^{56,57}\text{Ni}, t) = e_p(^{56,57}\text{Ni}) \, e^{-t/\tau_{56,57\text{Ni}}}$$

($\tau_{56,57\text{Ni}}$ are the mean lifetimes of $^{56}$Ni and $^{57}$Ni, and $t$ is the time after explosion), and $E_p(^{56,57}\text{Co}, t)$ are the energy production rates from decaying $^{56}$Co and $^{57}$Co

$$E_p(^{56,57}\text{Co}, t) = e_p(^{56,57}\text{Co})(e^{-t/\tau_{56,57\text{Co}}} - e^{-t/\tau_{56,57\text{Ni}}}).$$

For the transport calculations we use a frequency-independent opacity of $\chi_\gamma = \rho \cdot 0.027 \text{ cm}^2/\text{g}$ for the γ-photons (Swartz et al. 1995) and of $\chi_{e^+} = \rho \cdot 7.0 \text{ cm}^2/\text{g}$ for the positrons of the $^{56}$Co-decay (Colgate et al. 1997), whereas we assume a scattering atmosphere with a mean opacity of $\chi_{gray} = \rho \cdot 0.17 \text{ cm}^2/\text{g}$ for the optical photons. We note that the latter value affects mostly the rising branch of the light curve and the near-maximum phase (cf. Cappellaro et al. 1997), while the declining part of the light curve is primarily determined by $\chi_{e^+}$ and $\chi_\gamma$.

Figure 5 displays for different times after explosion the total energy deposition rates $E_{dep}(v, t)$ in erg/s/cm$^3$ as a function of depth.[9] As is shown, the γ-photons are hardly transported during the very early times after explosion and therefore the energy is deposited essentially only locally for the first week. Only after about seven days after explosion do the γ-photons reach the outer shells and a tiny fraction of about 0.02 % of their energy can escape from the atmosphere. After 18 days the γ-photons reach the iron-rich central part of the atmosphere, which contains almost no $^{56}$Ni, and after about 47 days the atmosphere becomes optically thin to the γ-photons and the deposition rate

---

[9] This simulation is based on the W7 explosion model (Nomoto et al. 1984).

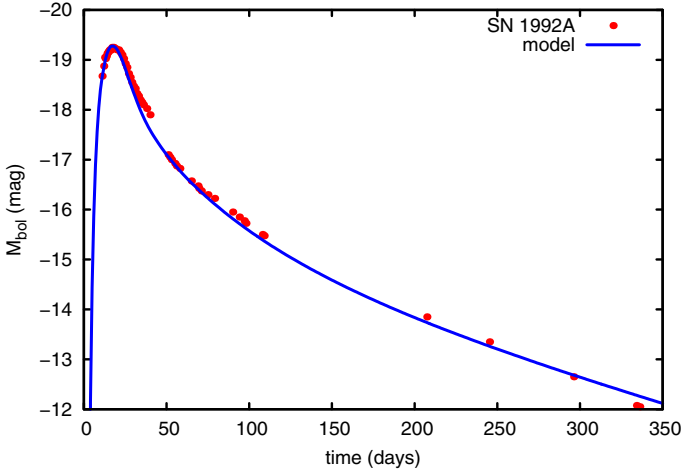

**Fig. 7.** Comparison of our calculated synthetic light curve (fully drawn) with the observed light curve of SN1992A (dots; Kirshner et al. 1993). The calculated light curve is based on the W7 explosion model.

is then proportional to the density stratification over the whole radius range.

For a point in time (25 days) closely after the brightness maximum and one (338 days) in the late, nebular phase we further display the individual energy deposition rates of $\gamma$-photons and positrons in Fig. 6. As shown, the $\gamma$-photons clearly dominate the energy deposition during the early phases, whereas in the late phases the light curve is driven mainly by the energy deposition of the positrons. This is of course because the mean free path lengths of the positrons are much smaller than those of the $\gamma$-photons, of which many escape during the late phases.

While in the time-dependent simulations the deposited energy is assumed to emerge as optical photons representing a mean radiation field,[10] in the non-LTE model for the calculation of the spectra the deposited energy is considered as an additional emissivity $\eta_\nu^{\text{dep}}$,

$$\eta_\nu^{\text{dep}}(v, t) = \frac{1}{4\pi} f_\nu(v, t) E_{\text{dep}}(v, t), \tag{1}$$

where $f_\nu$ is the normalized spectral distribution of the deposited energy (cf. Sect. 5.3).

To verify that our procedure describes the total energy deposition rates correctly we compare the synthetic light curve resulting from the time-dependent radiative transport to the observed light curve of SN Ia SN1992A (Kirshner et al. 1993) in Fig. 7. The comparison shows that this result of our simulation fits not only the observed steep increase in brightness during early times where the atmosphere is optically thick for all photons, but also the steep decrease after maximum brightness as well as the more shallow decrease from 50 to 300 days after explosion when the atmosphere becomes more and more optically thin. On basis of this encouraging result we regard our calculated deposition rates and luminosities derived from the time-dependent treatment as reliable quantities to be used for the calculations of synthetic SN Ia spectra.

## 4. Snapshots of the radiative transfer and synthetic spectra

In order to determine the physical properties of the most luminous stellar explosions via quantitative UV to IR spectroscopy another principal difficulty needs to be overcome: the diagnostic tools and techniques must be provided. This requires the construction of detailed atmospheric models and synthetic spectra for metal-dominated supernovae. The emitted spectra which are the basis of a quantitative analysis consist of thousands of strong and weak Doppler-shifted spectral lines forming a "pseudo-continuum", and an adequate and well-tested method is required to reproduce these structures accurately. Developing such a tool is not straightforward, since the corresponding simulations involve the solution of the statistical equilibrium for all important ions including the atomic physics (cf. Sect. 4.1), the radiative transfer equation at all transition frequencies, and the energy equation which represent a tightly interwoven mesh of physical processes (cf. Fig. 1). The most complicating effect in this system is the overlap of thousands of spectral lines of different ions, and with respect to this point the numerical method has to account not only for the blocking and blanketing influence[11] of all metal lines in the entire supersonically expanding atmosphere (cf. Pauldrach et al. 2001; Pauldrach et al. 2012), but also for a very special form of a deadlock of the lambda iteration which is induced by the mutual interaction of strong spectral lines (cf. Sect. 4.2).

Concerning the basic requirements for calculating detailed models for expanding atmospheres we have to concentrate on the following points:

- *The determination of the occupation numbers* via a solution of the *statistical equilibrium* containing collisional and radiative transition rates for all important ions as well as low-temperature dielectronic recombination. The rates are coupled to the hydrodynamic structure of the explosion model through the velocity field which enters into the radiative rates via the Doppler shift (as well as the combined effect of line blocking on the radiation field) and the density which affects the recombination and collisional rates.

- *The solution of the spherical radiative transfer equation* for the total opacities and source functions of all important ions and at every depth point, including the deeper "photospheric" layers where instead of the customary diffusion approximation an improved inner boundary for the radiation transfer under non-LTE conditions is used (cf. Sauer et al. 2006). (We note that the diffusion approximation is only applicable if the mean free paths of the photons are much shorter than any significant hydrodynamic length scale and the radiation field therefore thermalizes; unfortunately this condition is not fulfilled for SN Ia at and after maximum light, cf. Sauer et al. 2006). Our method uses an exact observer's frame solution (equivalent to a comoving frame solution) which correctly treats the angular variation of line opacities and emissivities. In this way the line profiles are not just spatially resolved, but multi-line effects and back-reactions of the line opacities on the model structures are also treated correctly.[12] Thus the method naturally takes into account the influence of the spectral lines – the strong UV *line blocking* including the effects of line-overlap[13] (cf. Fig. 8) – that affects the ionizing flux which determines the ionization and excitation of levels.

---

[10] This is a reasonable assumption, because the $\gamma$-photons generate fast electrons through Compton scattering which transfer the energy back to the radiation field either through Bremsstrahlung or ionizing processes (Sutherland & Wheeler 1984; Swartz et al. 1995).

[11] As we have found from previous model calculations that the behavior of most of the spectral lines depends critically on a detailed and consistent description of line blocking and line blanketing (cf. Pauldrach 1987; Pauldrach et al. 1990; Pauldrach et al. 1994; Pauldrach et al. 1996) special emphasis has to be given to a correct treatment of the Doppler-shifted line radiation transport and the corresponding coupling with the radiative rates.

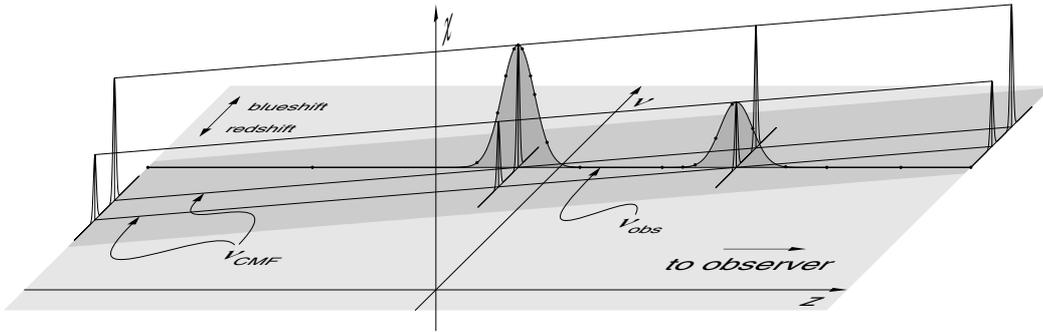

**Fig. 8.** Diagram illustrating the basic relationship of the rest-frame frequencies of spectral lines ($\nu_{CMF}$) to observer's frame frequency ($\nu_{obs}$) for one particular ray in the $z$ direction of the spherically symmetric geometry. Shown are two spectral lines of different opacity $\chi$ which get shifted across the observer's frame frequency by the (homologous) velocity field in the ejecta. The dots represent the stepping points of the adaptive microgrid used in solving the transfer equation in the radiative line transfer. The method employed is an integral formulation of the transfer equation using an adaptive stepping technique on every ray in which the radiation transfer in each micro-interval is solved as a weighted sum on the microgrid (cf. Pauldrach et al. 2001).

- *The determination of the temperature structure* via a solution of the *microscopic energy equation* (cf. Fig. 1) which, in principle, states that the energy – including the local energy production rates via the radioactive decay chains – must be conserved in the atmosphere. The radiation field, strongly influenced by line emission and absorption in the pseudo-continuum, clearly plays an important role (cf. *line blanketing*). The objective therefore is to calculate an atmospheric temperature stratification which conserves the total radiative flux and which treats the impact of the line opacities and emissivities properly.

  In general there are three methods for calculating electron temperatures in model atmospheres: (a) based on a condition of radiative equilibrium (this is modified by taking into account the energy source in the radiative transfer equation), (b) using a flux correction procedure (originating from Eq. 23 – cf. Fig. 1 and Sect. 5.3), and (c) based on the thermal balance of heating and cooling rates. Owing to the special physical circumstances in the atmospheres of SN Ia we use all three methods. The idea of the flux correction procedure is straightforward: the local temperature has to be adjusted in such a way that the total local radiative flux is conserved

---

[12] Since Jeffery (1993) and Dessart & Hillier (2005) found on basis of completely different models and procedures that relativistic effects are weak and only noticeable for the fastest observed velocities for the case of SNe, such effects have not been treated.

[13] If spectral lines of different opacity $\chi$ get shifted across an observer's frame frequency by the velocity field in the envelope, line-overlap, which is responsible for multiple scattering events, takes place (cf. Fig. 8). This problem is solved via an integral formulation of the transfer equation using an adaptive stepping technique on every ray in which the radiation transfer in each micro-interval is treated as a weighted sum on the microgrid:

$I(\tau_0(p,z)) = I(\tau_n)e^{-(\tau_n - \tau_0)} + \sum_{i=0}^{n-1}\left(e^{-(\tau_i - \tau_0)}\int_{\tau_i}^{\tau_{i+1}}S(\tau)e^{-(\tau-\tau_i)}\,d\tau(p,z)\right)$,

where $I$ is the specific intensity, $S$ is the source function and $\tau$ is the optical depth (increasing from $\tau_0$ on the right to $\tau_n$ on the left in Fig. 8). To accurately account for the variation of the line opacities and emissivities due to the Doppler shift, all line profile functions are evaluated correctly for the current microgrid-coordinates on the ray, thus effectively resolving individual line profiles (cf. Pauldrach et al. 2001).

(cf. Lucy 1964). This requires, however, that the temperature is the dominant parameter on which the flux depends, and that the effect of a change in temperature on the flux is known. In SN Ia ejecta these conditions are fulfilled only in the innermost regions before the density sharply decreases ($v < 15\,000$ km/s around maximum light). Thus, this method is applied as long as the condition of the modified radiative equilibrium (see Sect. 5.3) gives the same values of the temperature.

In the outer part of the expanding atmosphere, where scattering processes start to dominate and the effects of the line influence on the temperature structure are therefore more difficult to treat, we obtain the temperature structure from balancing energy gains and losses to the electron gas (heating and cooling rates), since this method is numerically advantageous for physical conditions where the opacity is dominated by scattering events that do not couple the radiation field to the thermal pool. In calculating the heating and cooling rates, all processes that affect the electron temperature have been included – bound-free transitions (ionization and recombination), free-free transitions, and inelastic collisions with ions. For the required iterative procedure we use a linearized Newton-Raphson method to extrapolate a temperature that balances the heating and cooling rates. With regard to these methods line blanketing effects which reflect the influence of line blocking on the temperature structure are naturally taken into account (cf. Pauldrach et al. 2001). Fig. 19 displays the resulting temperature structures for our best models.

We note that with the improvements to the iteration scheme we describe in this paper the final procedure yields a flux which is conserved to a few percent (e.g., Fig. 15). Without these improvements the models showed a severe flux loss of up to 50% – a first hint that a general problem exists in the lambda iteration under SN Ia conditions.

In principle, the iterative solution of the total system of equations then yields the synthetic spectrum. However, as will be discussed below (Sect. 4.2), the convergence of a SN Ia model requires more than just a straightforward implementation of the principal features of the method just described.

### 4.1. Atomic models

Ionization and excitation play the major role in calculating the spectrum of a SN Ia as the ionization balance leaves its imprint on the spectra via the strength and structure of the spectral lines formed throughout the envelope. Since almost all of the ionization thresholds lie within the UV spectral range which is shaped by thousands of not directly observable spectral lines, the computed ionization balances of all elements depend ultimately on the quality of the calculated occupation numbers of the spectral lines which produce the "pseudo-continuum" of ionizing radiation. These in turn are directly dependent on the quality of the atomic data, and the atomic data are – from this point of view – therefore observable quantities (even the atomic data of elements which have lines in the UV but not the visible part of the spectrum). With regard to this essential point the pitfall of "garbage in, garbage out"[14] must of course be avoided, and we have thus revised and improved the basis of our model calculations, the atomic models, extensively.

**Table 2.** Summary of revised atomic data. In columns 2 and 3 the number of levels used in the non-LTE calculations are given in packed and unpacked form. Columns 4 and 5 list the number of lines used for solving the rate equations (the statistical equilibrium for all important ions) and for the line blocking and synthetic spectra calculations, respectively.

| Ion | levels packed | unpacked | lines rate eq. | blocking | Ion | levels packed | unpacked | lines rate eq. | blocking |
|---|---|---|---|---|---|---|---|---|---|
| C I   | 22 | 39 | 101 | 1397  | Ti I   | 11 | 14 | 15 | 2813 |
| C II  | 36 | 73 | 284 | 11079 | Ti II  | 50 | 112 | 509 | 56695 |
| C III | 50 | 90 | 520 | 4406  | Ti III | 50 | 119 | 482 | 8602 |
| C IV  | 27 | 48 | 103 | 24    | Ti IV  | 12 | 18 | 18 | 48 |
| O I   | 48 | 85 | 537 | 18326 | Cr I   | 12 | 12 | 6 | 4105 |
| O II  | 50 | 117 | 598 | 39215 | Cr II  | 50 | 135 | 321 | 221973 |
| O III | 50 | 102 | 554 | 24514 | Cr III | 50 | 141 | 447 | 134711 |
| O IV  | 44 | 90 | 435 | 17933 | Cr IV  | 50 | 117 | 486 | 60264 |
| Mg I   | 21 | 34 | 74 | 483    | Mn I   | 9 | 9 | 7 | 2088 |
| Mg II  | 26 | 41 | 104 | 3373  | Mn II  | 50 | 129 | 305 | 227824 |
| Mg III | 50 | 96 | 529 | 2457  | Mn III | 50 | 133 | 364 | 175645 |
| Mg IV  | 50 | 117 | 589 | 3669 | Mn IV  | 50 | 124 | 467 | 131822 |
| Si I   | 21 | 27 | 90 | 1420   | Fe I   | 50 | 137 | 557 | 822167 |
| Si II  | 40 | 59 | 293 | 4560  | Fe II  | 50 | 148 | 452 | 849612 |
| Si III | 50 | 90 | 480 | 4044  | Fe III | 50 | 126 | 263 | 756406 |
| Si IV  | 25 | 45 | 90 | 245    | Fe IV  | 45 | 126 | 253 | 172930 |
| S I   | 17 | 22 | 32 | 310     | Co I   | 50 | 121 | 500 | 284033 |
| S II  | 50 | 87 | 595 | 27775  | Co II  | 50 | 134 | 482 | 189911 |
| S III | 14 | 21 | 35 | 196     | Co III | 50 | 141 | 469 | 200658 |
| S VI  | 50 | 107 | 547 | 4547   | Co IV  | 41 | 97 | 70 | 146252 |
| Ar I   | 31 | 31 | 141 | 970    | Ni I   | 42 | 42 | 212 | 1904 |
| Ar II  | 16 | 35 | 48 | 1979   | Ni II  | 50 | 114 | 439 | 129978 |
| Ar III | 13 | 31 | 24 | 404    | Ni III | 40 | 102 | 281 | 131508 |
| Ar VI  | 11 | 11 | 25 | 158    | Ni IV  | 50 | 146 | 528 | 183367 |
| Ca I   | 50 | 132 | 479 | 6992   | Cu I   | 49 | 98 | 329 | 23641 |
| Ca II  | 28 | 28 | 126 | 3067   | Cu II  | 50 | 121 | 513 | 170516 |
| Ca III | 15 | 15 | 43 | 638    | Cu III | 13 | 21 | 41 | 490 |
| Ca IV  | 50 | 129 | 176 | 20031 | Cu IV  | 50 | 124 | 477 | 17466 |

Up to now the atomic models of all of the important ions of the 149 ionization stages of the 26 elements considered (H to Zn, apart from Li, Be, B, and Sc) have been revised in order to improve the quality. This has primarily been done using the Autostructure program (Eissner et al. 1974; Nussbaumer & Storey 1978), which employs the configuration-interaction approximation to determine wave functions and radiative data. The improvements include energy levels (comprising a total of about 5,000 observed levels, where the fine structure levels have been "packed" together[15]) and transitions (comprising more than 33 000 bound-bound transitions for the non-LTE calculations and more than 6 500 000 lines for the blocking calculations[16], and 20,000 individual transition probabilities of low-temperature dielectronic recombination and autoionization). Additional line data were taken from the line list of Kurucz (1992) and Kurucz (2007): approximately 50 000 lines have been added to the Autostructure data for ions of Mn, Fe, Co, and Ni. These concern transitions to even higher levels than those having been calculated with Autostructure, but which might nonetheless be of significance in the blocking calculations. We also added forbidden line transitions from Aller (1984) for He, C, N, O, Ne, Na, Mg, Si, S, Cl, Ar, K, Fe, and Ni in case they were missing. From the Opacity Project (cf. Seaton et al. 1994; Cunto & Mendoza 1992) another 4 466 lines have been included, as well as photoionization cross-sections (almost 2 000 data sets have been incorporated). Collisional data have become available through the IRON project (see Hummer et al. 1993) — almost 1 300 data sets have been included. Table 2 gives an overview of the ions affected by the improvements, and Fig. 9 show two representative Grotrian diagrams of the line transitions – for the low ionized intermediate mass element S II, and the low ionized iron-group element Fe II – as used in the statistical equilibrium of the non-LTE model.

---

[14] In memoriam Anne Underhill.

[15] Note that artificial emission lines may occur in the blocking calculations if the lower levels of a fine structure multiplet are left unpacked but the upper levels of the considered lines are packed.

[16] The Autostructure calculations involve many more excited levels than actually used in the explicit non-LTE rate equation system; our line list does, however, include transitions to such highly excited levels above our limit of considering the level structure – occupation numbers of these upper levels are calculated using the two-level approximation (augmented with the acceleration term of Eq. 18) on the basis of the correctly calculated occupation number of the lower level.

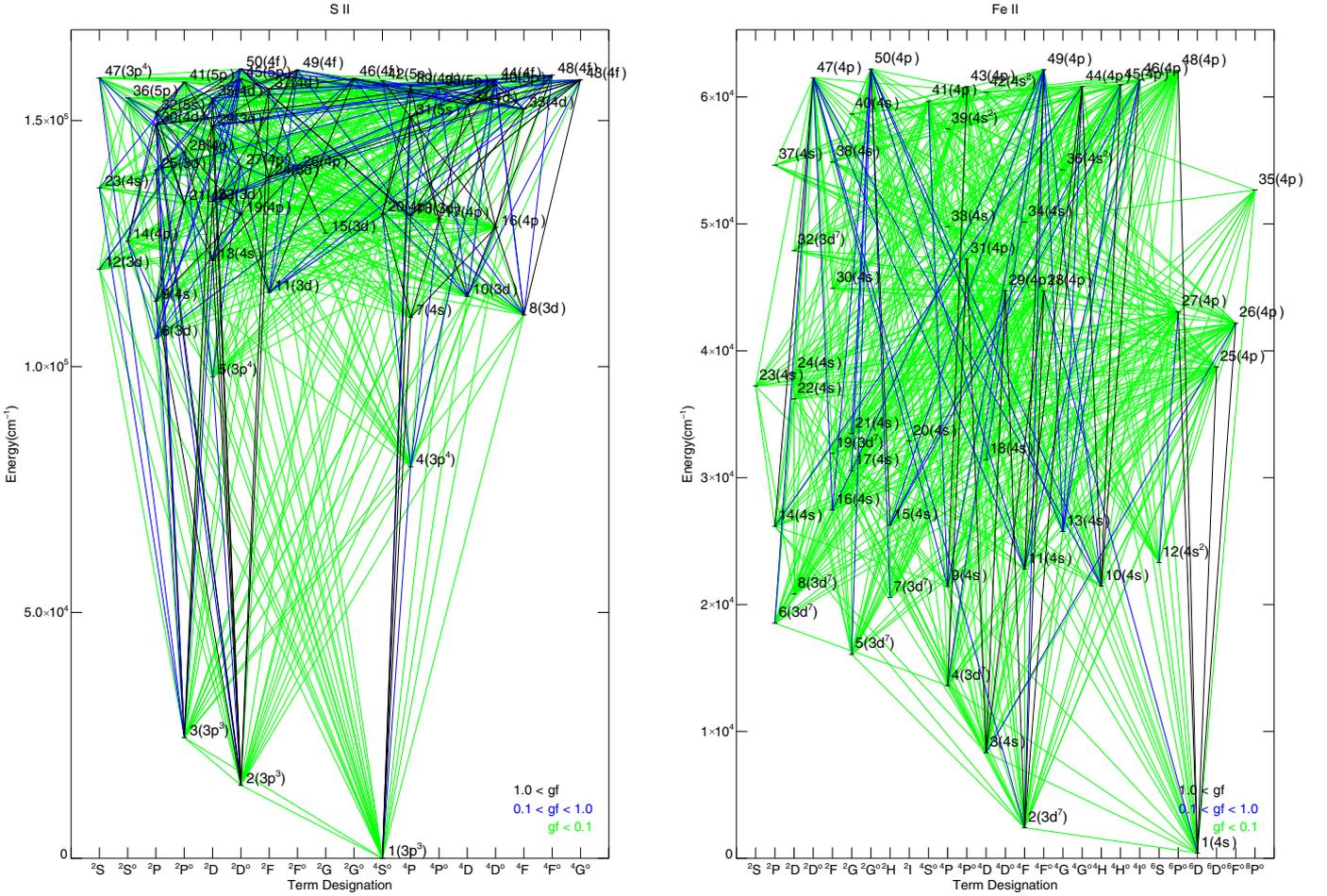

**Fig. 9.** Grotrian diagram of S II (left) and Fe II (right) as used in the statistical equilibrium of the non-LTE model (note that our line list does also include all transitions which are connected via their lower level to the level structure shown – these lines additionally are taken into account for the blocking calculations (cf. Table 2).)

### 4.2. The ansatz of an accelerated lambda iteration procedure for the mutual interaction of strong spectral lines (ALImI)

Improvements of our method regard not just the inclusion of the energy deposition and the modifications of the atomic database, but also a sophisticated treatment of the strong spectral lines which interact in an interwoven way with a "pseudo-continuum" entirely formed by these Doppler-shifted spectral lines[17]. An ap-propriate treatment of this effect is obviously of great importance for the computation of synthetic SN Ia spectra, because the blocking is most effective in the wavelength region where the radiation emitted from the photosphere reaches its peak intensity (the temperatures of the outer photospheric region are typically of the order of 10 000 K near maximum light). The high velocities complicate the problem even more, since the number of spectral lines which the Doppler-shift brings into resonance at different depths in the envelope is very large. Thus, the effect of line blocking is to reduce the intensity of the radiation field in the UV, and because of the low photospheric temperatures of SN Ia, which leads to an ionization regime of low stages which have an enormous number of line transitions in the corresponding frequency region (cf. Fig. 10), this reduction is strong in the UV. But line blocking has an effect not only on the emergent spectra, but also indirectly on the level populations, since the rate of ionization from excited levels occurring at wavelengths correspond-cesses.) The number of such lines is very large, since the envelope matter reaches velocities on the order of 20 000 km/s and in the UV, where the metal lines crowd very densely, this line overlap becomes a quasi-continuous source of opacity (cf. Fig. 10). A UV photon can thus only escape if during its random walk being scattered and re-scattered in the optically thick atmosphere it manages to connect to the thermal pool, is reborn as a redder photon, and finds a low-opacity window in the spectrum (cf. Pauldrach et al. 1996).

---

[17] Due to the large velocity gradients in the envelope (homologous expansion, $v \propto r$) the UV part of the spectrum is effectively blocked by the superposition of thousands of strong metal absorption lines within the expanding envelope, since a photon emitted at the photosphere will be redshifted in the co-moving frame as it moves outwards in the envelope, and will be brought into resonance with redder and redder lines. (For the composition of the ejecta the redmost important ground state ionization edge which contributes to *continuous* absorption belongs to Ca II at 1044Å; if not for the spectral lines, the frequencies redward of this edge would otherwise be almost free from any absorption pro-

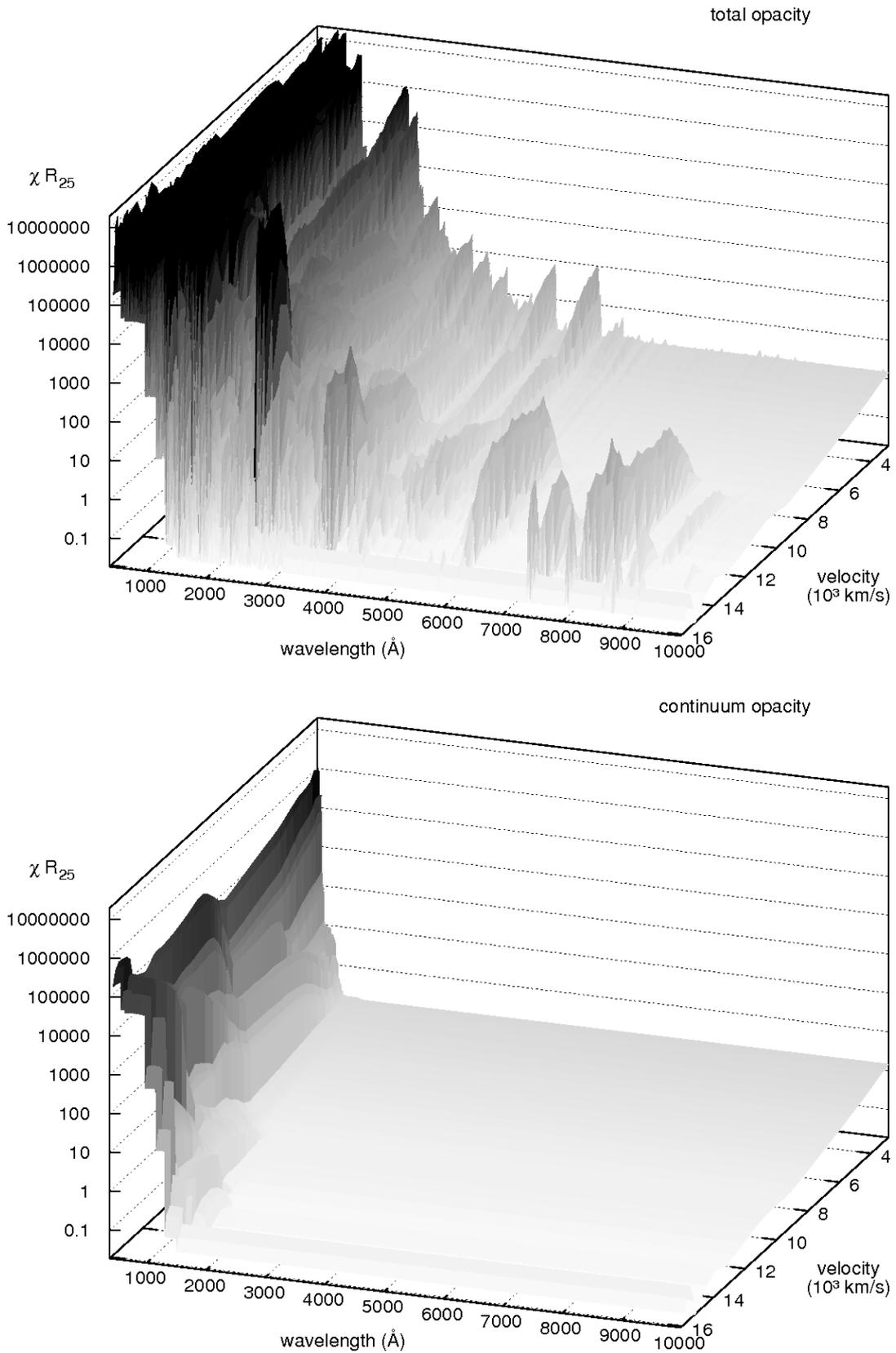

**Fig. 10.** Logarithm of the total opacity (upper diagram) and the true continuum plus Thomson scattering opacity (lower diagram) times the radius $R_{25}$ (defined for a velocity of $v = 3\,500$ km/s and a time at day 25) versus velocity and wavelength for a SN Ia model at early phases. Because of the small absolute densities (compared to stars) and a composition which is dominated by low ionization stages of intermediate-mass and iron-group elements, a weak free-free and bound-free continuum is obtained in the optical and infrared part of the spectrum. Therefore, electron scattering opacity becomes the dominating source of opacity at wavelengths redward of about 5000 Å, even for deeper layers of the envelope, whereas the total continuum opacity is completely irrelevant when compared to the line opacity blueward of about 4000 Å. This behavior illustrates the formation of the "pseudo-continuum" in the envelopes of SN Ia, which just results from the overlap of the thousands of spectral lines shown.

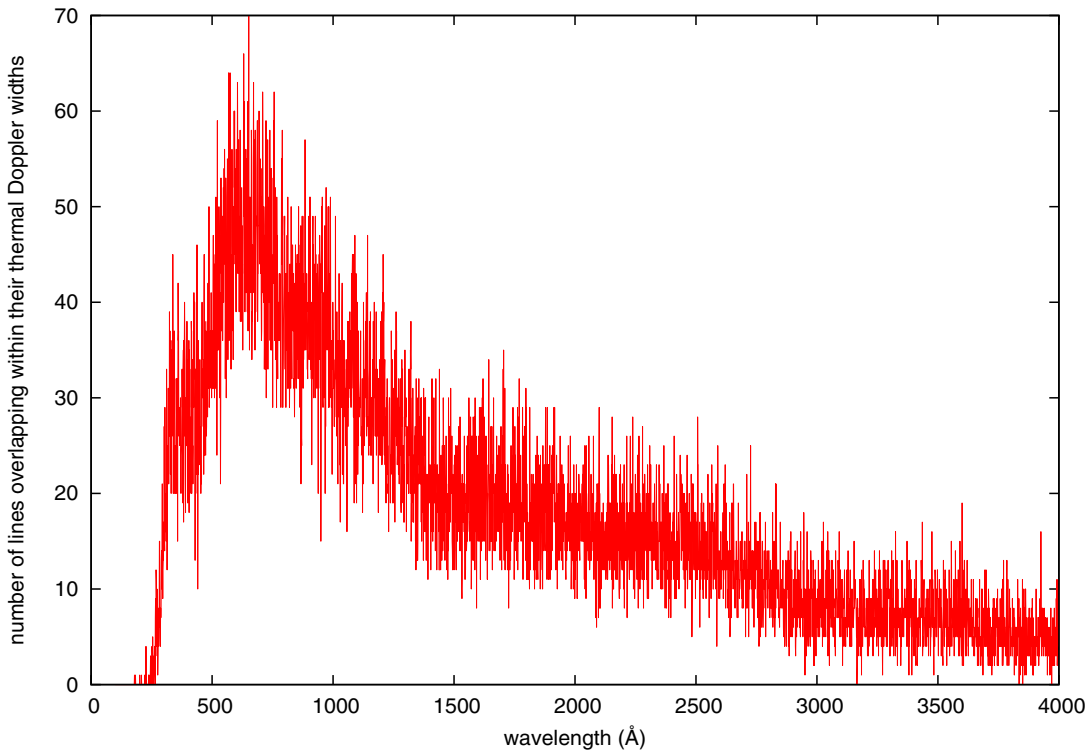

**Fig. 11.** Number of lines overlapping within their thermal Doppler widths $\Delta\nu_D = \sqrt{2kT/m_{ion}} \cdot \nu_0/c$ as a function of wavelength. (Only lines from the lowest three ionization stages, excluding forbidden lines, were considered for this plot. The overwhelming majority of these lines is from iron-group elements. The thermal Doppler width was calculated for a representative element of this element-group using a temperature typical for SN Ia, and is on the order of 0.01 Å at a wavelength of 1000 Å.) The behavior that up to 70 lines from different elements of the iron-group dominated SN Ia ejecta overlap per Doppler width is the origin of a hidden form of deadlock of the lambda iteration used to solve the multi-level non-LTE problem, and a straightforward lambda iteration will fail to converge under such circumstances unless special measures are taken. *This lambda iteration deadlock is a general problem because it is solely based on the fact that a large number of lines overlap within the thermal Doppler widths of the lines*, and this behavior cannot be influenced by choosing extremely narrow frequency and radius grids, or by choosing a different procedure to solve the radiative transfer (comoving frame or observer's frame).

ing to the line-blocked region is, for instance, reduced. Thus, an iteration scheme which includes the effects of line blocking is required in order to compute the level populations and the radiation field consistently.

Although we previously thought that this problem had already been solved in the past, it now seems that the real problem – produced by the iteration scheme which connects the mutual interaction of strong spectral lines with the "pseudo-continuum" (itself entirely formed by these spectral lines), the level populations, and the radiation field – had been overlooked up to now, since we had not calculated deep enough into the photosphere. But this is required, because most of the synthetic spectra calculated up to now for the early epochs show, compared to observations, excess flux in the red and infrared wavelength regions (cf. Pauldrach et al. 1996, Nugent et al. 1997, Sauer et al. 2006 and the spectral fits of Stehle et al. 2005). This is the case because most models assume (directly or indirectly) thermalization at the inner boundary. But, compared to a blackbody spectrum at the respective temperature of the usually applied inner boundary, the radiation field in this region is more likely to show a bluer characteristic, because the opacity distribution (cf. Fig. 10) does not lead to a considerable optical thickness (which is required for thermalization) in this frequency range, and because the radiation resulting from the deposition of $\gamma$ rays is only partly thermalized (cf. Sect. 3) – emission from the down-scattering of $\gamma$-photons cannot be generated farther out in the ejecta in wavelength ranges that do not exhibit significant continuum or line opacities. Thus, the characteristic shape in the red and infrared wavelengths ranges of SN Ia spectra at early epochs must show a slope in the continuum which is generally steeper than the slope of a corresponding blackbody spectrum, and this is of course the case for the observed spectra. To obtain the same behavior for the synthetic spectra one has to calculate much deeper into the photosphere (cf. Sect. 5.1), but this produces a non-trivial problem to the iteration scheme of the level populations and the radiation field due to the extreme optical depths reached in the UV. Concerning this point we have made significant progress in analyzing the complicated process involved, and developing a suitable procedure to overcome a special form of deadlock of the lambda iteration connected to this iteration process.

**Stagnation of the lambda iteration due to the mutual interaction of strong spectral lines.** If we calculate deep into the photosphere, the temperature increases significantly (cf. Fig. 19) and the ionization balance therefore shifts to higher stages in accordance with the radiative flux maximum which is to be found in the range of the UV (cf. Fig. 15). As the spectral lines of the higher ionization stages are primarily also located in the frequency regime of the UV (cf. Fig. 10), this flux is blocked. Although this appears to be a dead end, the radiation has to be transported through the envelope. Thus, the UV photons have to connect to the thermal pool as they scatter around, become redder photons, up to the point they find a low-opacity window in the spectrum, and escape from the photosphere. In stellar atmospheres, much of this is accomplished by bound-free and free-free processes (the true continuum). In SN Ia, however, these true continuum opacities are much too small (cf. Fig. 10) to effectively realize this mechanism. The chain of required processes – starting with the collisions, which connect the occupation numbers to the thermal pool, and proceeding with the true continuum opacities and emissivities defined by those occupation numbers, which can absorb at high energies and emit at lower energies, thus transforming UV photons to redder photons – can therefore not take place directly (as in the atmospheres of stars), and must thus be realized in the atmospheres of SNe Ia in a different way.

The origin of the problem is rooted in the high spectral line density of the UV spectral range where a small wavelength interval of only 1 Å can contain as many as 7000 lines from different elements of the iron-group dominated SN Ia ejecta, and this means that within the even much smaller wavelength ranges of the thermal Doppler widths of the lines – on the order of 0.01 Å at a wavelength of 1000 Å – up to seventy lines from different elements can overlap (cf. Fig. 11). This extreme behavior is exacerbated by the fact that up to 20% of these lines have a line strength that leads to an optical depth[18] of $\tau_{line} > 1$ (cf. Fig. 10).

Thus, strong spectral lines are not only the origin of the pseudo-continuum of the atmospheres of SNe Ia, but their mutual interactions are the reason for a compounded form of line blocking with which a stagnation of the lambda iteration (see below) is inherently connected. As such a slow-down can inhibit the lambda iteration from converging to the true selfconsistent solution (cf. Mihalas 1978), it has to be explicitly accounted for in order to describe the source function along with the escape of the blocked UV photons via the radiative transfer correctly. It is further important to realize that the deadlock of the lambda iteration we describe here is *a general problem*. This is because *a large number of lines from different elements overlap even within the small wavelength ranges of their thermal Doppler widths*, and this means that the problem is independent of the frequency and depth grid spacing used in the model, and also independent of whether the radiative transfer is solved in a frame comoving with the fluid or at rest with the observer. But it also means that the problem only appears if the model considers a certain minimum number of lines in the comprehensive non-LTE iteration scheme, such as the millions of atomic line transitions required for a realistic solution of supernova ejecta.

To uncover this hidden form of a deadlocked lambda iteration for the first time we start with the general solution of the radiative transfer

$$J_\nu(r) = \Lambda(S_\nu(r)), \qquad (2)$$

where the frequential mean intensity $J_\nu(r)$ is described by the so-called $\Lambda$-operator, which mathematically represents the radiation field as a function of the source functions $S_\nu(r)$ calculated previously from the occupation numbers obtained from the statistical equilibrium (cf. Fig. 1). The naive way of obtaining a selfconsistent solution of radiation field and occupation numbers is the straightforward $\Lambda$-iteration, where the radiative transfer is solved at a considered frequency $\nu_r$ alternately with the source functions $S_m(r) = \eta_m(\nu, \mu, r)/\chi_m(\nu, \mu, r)$ obtained for all specific lines $m$ from the opacities $\chi_m(\nu, \mu, r)$ and emissivities $\eta_m(\nu, \mu, r)$

$$\chi_m(\nu, \mu, r) = n_j\left(\frac{n_i}{n_j} - \left(\frac{n_i}{n_j}\right)^* e^{-h\nu/kT}\right)\alpha_m(\nu, \mu, r)$$
$$\eta_m(\nu, \mu, r) = \frac{2h\nu^3}{c^2} n_j \left(\frac{n_i}{n_j}\right)^* e^{-h\nu/kT} \alpha_m(\nu, \mu, r), \qquad (3)$$

---

[18] The optical depth scale is defined by the integral over the (Doppler-shifted) line opacity $\chi_{line}$ along a ray: $\tau_{line} = \int_0^\infty \chi_{line}(z)\,dz$. In SN Ia, due to the homologous expansion, this integral is independent of angle.

where

$$\alpha_m(\nu, \mu, r) = \frac{\pi e^2}{m_e c} f_m \phi_m\left(\nu - \nu_m\left(1 + \mu\frac{v(r)}{c}\right)\right) \qquad (4)$$

are the radiative cross sections.[19] The radiation field enters into the radiative transition rates

$$R_{ij}(r) = 2\pi \int\limits_0^\infty \int\limits_{-1}^{+1} \frac{\alpha_m(\nu, \mu, r)}{h\nu} I_\nu(r, \mu)\,d\mu\,d\nu = \frac{4\pi}{h\nu_m}\frac{\pi e^2}{m_e c} f_m \overline{J}_m(r)$$

$$R_{ji}(r) = 2\pi\left(\frac{n_i}{n_j}\right)^* \int\limits_0^\infty \int\limits_{-1}^{+1} \frac{\alpha_m(\nu, \mu, r)}{h\nu}\left(\frac{2h\nu^3}{c^2} + I_\nu(r, \mu)\right) e^{-h\nu/kT}\,d\mu\,d\nu, \qquad (5)$$

(where $I_\nu(r, \mu)$ is the specific intensity at frequency $\nu$ and

$$\overline{J}_m(r) = \frac{1}{2}\int\limits_0^\infty \int\limits_{-1}^{+1} \phi_m(\nu, \mu, r) I_\nu(r, \mu)\,d\mu\,d\nu \qquad (6)$$

is the mean intensity in line $m$) that are used to determine the occupation numbers. Because these in turn are determine the opacities and emissivities for the radiative transfer

$$\overline{J}_m(r) = \Lambda(S_m(r), S_{cont}(r)), \qquad (7)$$

this self-referential procedure does not automatically converge to the correct values. Indeed, it is well known that this procedure does not correctly converge for optically thick cases[20].

A technique to overcome this problem is the complete-linearization method of Auer & Mihalas (1969), but as this technique is founded on a multidimensional Newton-Raphson

---

[19] We use the standard conventions for the notation: $c$ is the speed of light, $v(r)$ is the local velocity, $T(r)$ is the local temperature of the material, $e$ is the charge and $m_e$ the mass of an electron, $f_m$ is the oscillator strength of a line with index $m$, rest frequency $\nu_m$, and local occupation numbers $n_i(r)$ and $n_j(r)$, $\phi_m(\nu, \mu, r)$ is the line profile function accounting for Doppler-shifting of the lines along the ray, $(n_i/n_j)^*$ is the Saha-Boltzmann factor, and $\mu$ is the cosine of the angle between the ray direction and the outward normal on the spherical surface element.

[20] The radiative rate coefficients for bound-bound (line) transitions in terms of the Einstein coefficients $A_{ji}$, $B_{ji}$, and $B_{ij}$ are given by

$$R_{ij}(r) = \overline{J}(r) B_{ij}$$
$$R_{ji}(r) = \overline{J}(r) B_{ji} + A_{ji}$$

where $\overline{J}(r)$, the mean intensity in the line, can be computed using the comoving-frame or the Sobolev-with-continuum method. $\overline{J}(r)$ contains contributions from the ambient radiation field $I_\nu(r, \mu)$ as well as from the line source function $S_m(r)$ itself. If the line is optically thick, this latter contribution will dominate, so that a straightforward iteration scheme to find a consistent solution for the radiation field and the occupation numbers by means of the $\Lambda$-iteration

$$n_i^{(old)},\ n_j^{(old)} \quad \rightarrow \quad S_m^{(old)}(r) = \frac{n_j^{(old)}(r) A_{ji}}{n_i^{(old)}(r) B_{ij} - n_j^{(old)}(r) B_{ji}}$$

$$\Lambda(S_\nu^{(old)}(r)) \quad \rightarrow \quad I_\nu(r, \mu)$$

$$I_\nu(r, \mu),\ S_m^{(old)}(r) \quad \rightarrow \quad \overline{J}(r) \quad \rightarrow \quad R_{ji}(r), R_{ij}(r) \quad \rightarrow \quad n_i^{(new)}, n_j^{(new)}$$

will converge only slowly, if at all, since in the optically thick case the dominant line source function tends to sustain itself in this scheme even if it does not have the correct consistent value.

method, the description of the model atoms used is necessarily limited. Based on the ideas of Cannon (1973), Scharmer (1981), Scharmer & Carlsson (1985), Werner & Husfeld (1985), Olson et al. (1986), and Pauldrach & Herrero (1988) operator perturbation methods have therefore been developed which overcome this weakness even when applied to multi-level non-LTE calculations. However, these accelerated lambda iteration (ALI) procedures solve the problem of a not correctly converging lambda iteration only for cases where at every chosen frequency and radius point only a single process dominates the opacity[21].

But this is not the case for the radiative transfer in the SNe Ia envelopes, since there are no significant continuum sources over a large frequency range in these ejecta; instead millions of strong lines overlap to form the pseudo-continuum. *At any frequency and depth point the radiation field is therefore influenced by several tens of strong lines of different ions, and the classical accelerated lambda iteration as well as the common complete-linearization approach* (the simplest form of an accelerated lambda iteration) *simply fail*, because they are based on the premise that at each frequency and depth point only the one dominant process that may have "frozen" the iteration must be accelerated to let the system converge (see also Sect. 5.2).

As Mihalas (1978) has already described comprehensively in his book, the stagnation of the lambda iteration is one of the fundamental physical problems inherent in the solution of the non-LTE problem, and the failure of a lambda iteration to converge is a pitfall of crucial importance for non-LTE radiative transfer models. Under SN Ia conditions with mutually interacting strong spectral lines we encounter the general problem of non-convergence again, since although the model correctly implements the known ALI procedures for the continuum and optically thick single lines, these fail to solve the problem of multiple opacities belonging to different physical transitions acting together at a given frequency and radius point. Because in SN Ia strong spectral lines are present at almost all radius points at almost all frequencies in the UV, the mean intensity in a line at a given radius point contains not only contributions of that line

---

[21] The accelerated lambda iteration involves analytically canceling the contribution of the line itself in the formulation of the rate equations describing the statistical equilibrium. This is particularly easy to demonstrate in Sobolev theory, where the line contribution becomes a purely local term,

$$\overline{J}(r) = (P_{\mathrm{I}} + S_{\mathrm{cont}}\overline{U}) + (1 - P_{\mathrm{S}} - \overline{U})S_m(r)$$

(here, $P_{\mathrm{I}}$ describes the interaction of the line with the incident intensity, $P_{\mathrm{S}}$ is the probability for a line photon to escape the resonance zone of the line, and $\overline{U}$ describes the interaction between line and the background continuum within the resonance zone, cf. Puls & Hummer 1988.) If we write the net radiative rate $n_i^{(\mathrm{new})}R_{ij} - n_j^{(\mathrm{new})}R_{ji}$ of the transition using the rate coefficients and the mean line intensity with $S_m(r)$ expressed in terms of the "new" occupation numbers, then, after canceling the appropriate terms, we see that the transition is identically described by the rate coefficients

$$\tilde{R}_{ij} = (P_{\mathrm{I}} + S_{\mathrm{cont}}\overline{U})B_{ij}$$
$$\tilde{R}_{ji} = (P_{\mathrm{I}} + S_{\mathrm{cont}}\overline{U})B_{ji} + (P_{\mathrm{S}} + \overline{U})A_{ji}$$

which do not anymore directly depend on the "old" source function of the line. Similar options exist for accelerating the convergence in optically thick bound-free (continuum) transitions (e.g., Pauldrach & Herrero 1988). These methods have been very successful in stellar atmosphere modeling, where, although their combined influence can lead to a substantial blocking of radiation, individual lines are nevertheless fairly isolated, and at any frequency there is usually just one dominant contributor to the continuum opacity.

itself and the continuum, but also contributions of all other lines at that frequency $\nu_l$ and radius point $r$. The influence of multiple intermediate-strength opacities at a given point is still a problem, because although each single intermediate opacity would converge in accordance with an uncritical optically thin behavior, the total influence of the multiple intermediate opacities still maintains the behavior of a critical optically thick case.

To derive the modification required to correctly converge the iterative procedure which determines the mean intensity in a line we begin by replacing Eq. 7 by

$$\overline{J}_m(r) = \Lambda(S_1(r), S_2(r), S_3(r), \ldots, S_m(r), \ldots, S_{\mathrm{cont}}(r)), \quad (8)$$

which can also be written as

$$\overline{J}_m(r) = \Lambda(S_{\mathrm{L}}(r, \mu), S_{\mathrm{cont}}(r)), \quad (9)$$

where $S_{\mathrm{L}}$ is the total line source function given by

$$S_{\mathrm{L}}(r, \mu) = \sum_m \frac{\chi_m(\nu_l, \mu, r)}{\chi_{\mathrm{L}}(\nu_l, \mu, r)} S_m(r), \quad (10)$$

and where $\chi_{\mathrm{L}}(\nu_l, \mu, r)$ is simply the total line opacity at each considered frequency

$$\chi_{\mathrm{L}}(\nu_l, \mu, r) = \sum_m \chi_m(\nu_l, \mu, r). \quad (11)$$

In the comoving frame where the line opacities are isotropic and the line profile functions are given by

$$\phi_m(\nu_l, \mu, r) = \phi(\nu_l - \nu_m, r) = \frac{1}{\sqrt{\pi}\Delta\nu_{\mathrm{D}}(r)} e^{-(\nu_l - \nu_m)^2/\Delta\nu_{\mathrm{D}}^2(r)},$$

the above summations must run over all lines $m$ that lie within a few Doppler widths of the considered line's frequency. In the observer's frame, all lines whose maximum Doppler shift $\Delta\nu = \pm \nu_m v_{\mathrm{max}}/c$ puts them in range of a given observer's frame frequency can in principle affect the radiation field at that frequency. In Fig. 8, these correspond to those lines whose rest frequencies lie in the gray band in the $(\nu, z)$-plane at $z = 0$. In our procedure, for every depth point and all atomic species, the correct Doppler-shifted thermal broadening based on atomic weight and local temperature is then used to check whether the individual line profile functions overlap in the radius interval around that grid point.

In the frame of an operator perturbation technique Eq. 9 can in principle be written in the form

$$\overline{J}_m(r) = \Lambda(S_{\mathrm{L}}(r, \mu), S_{\mathrm{cont}}(r)) + \Lambda^* \left(S_{\mathrm{L}}^{(i)}(r) - S_{\mathrm{L}}^{(i-1)}(r)\right) \quad (12)$$

(cf. Pauldrach & Herrero 1988). The basic idea of such an approach is that the radiation field affecting the rate equations in the current iteration ($i$) is obtained from a combination of the stabilizing properties of the $\Lambda$-iteration and the application of an approximate operator $\Lambda^*$ which acts on the inconsistency of the total line source function of the current iteration ($i$) and the total line source function of the previous iteration ($i - 1$), with the constraint that the corrective term goes to zero when true convergence is achieved. This technique basically defines the principle of the ALImI method (accelerated lambda iteration for mutual interactions) to be used. (With respect to the approximate term $\Lambda^*$ we take for the directional component of the total line source functions that of the central ray as being representative for all other rays, evaluated at the frequency of the line considered.)

Equations 5–12 together describe the general ansatz of an accelerated lambda iteration procedure needed to solve the problem of mutual interaction of strong spectral lines. However, this system does not yet represent a workable solution to the problem because the mode of operation by which an approximate lambda operator $\Lambda^*$ may accelerate the convergence for each transition in every interacting group of lines remains to be specified. The methods that have already been developed to solve the problem for single spectral lines are not applicable for this more complex case. (Neither can the common complete-linearization approach that treats each element separately address this problem when the interaction involves lines from different elements.)

Up to now we have therefore just *uncovered a hidden form of a lambda iteration deadlock*, and we have indicated that a more sophisticated treatment of the line processes involved in the simulations of expanding atmospheres of supernovae Ia is required to obtain convergence of the NLTE-model and thereby true consistency of occupation numbers and radiation field. Of course we expect that this more consistent description should lead to changes in the energy distributions and line spectra which in turn should lead to a much better agreement with the observed spectra than previous models. As this will obviously have important repercussions for quantitative analysis of metal-dominated supernovae spectra, we will investigate its behavior in the next section, and we will also discuss how the problem of the mutual interaction of strong spectral lines can be solved.

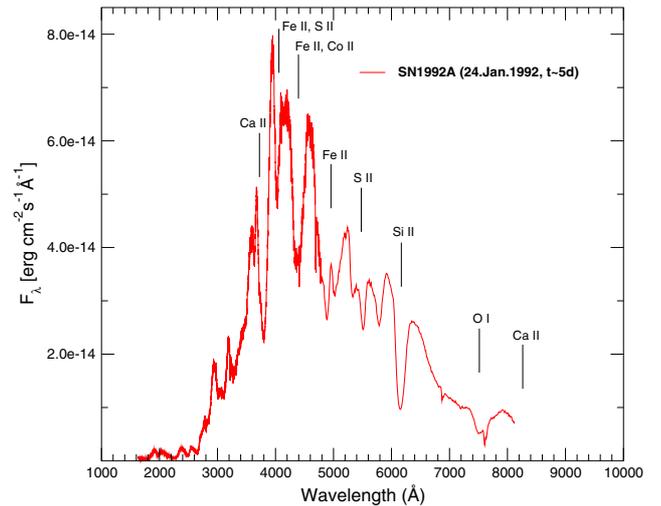

**Fig. 12.** Observed HST spectrum of a standard supernova of type Ia at early phase – SN 1992A approximately 5 days after maximum, representing an epoch of about 25 days after explosion (Kirshner et al. 1993). Since Kirshner et al. (1993) have found no reason to correct for extinction we have taken their data as-is. Furthermore, we are only interested in the shape of the spectrum, not the (distance-dependent) absolute flux, so we will treat the scaling as a free parameter in the following comparison plots.

## 5. Model calculations and discussion

The question whether the spectral energy distributions calculated from our current models of SN Ia are already realistic enough to be used for diagnostic issues of course requires a test, which can only be provided by a comparison of the synthetic and observed spectra of individual objects at different epochs. In this paper we will, however, restrict ourselves to models representing a generic "normal" SN Ia around maximum light (approximately 25 days after explosion), and we base our model calculations on the venerable W7 explosion model of Nomoto et al. (1984). As observational counterpart we have selected SN 1992A (Kirshner et al. 1993) which has a good coverage of the blue part of the spectrum with a reasonable S/N.

Such comparisons are important tests, since they involve not only hundreds of directly observable spectral signatures of various ionization stages with different ionization thresholds covering a large frequency range, but also thousands of not directly observable spectral lines which influence the ionization balance via the ionizing flux. But even this latter influence can be traced by the spectral lines in the observable part of the spectrum, since almost all of the important ionization thresholds lie within the spectral range of the UV and are therefore influenced by the lines located in that spectral region. The quality of the spectral energy distribution thus stands on the quality of the calculated ionization balance, and the synthetic spectrum, which is just a function of the basic parameters and the explosion model, is therefore as realistic (or unrealistic) as the simulation itself.

### 5.1. First results from a comparison of synthetic and observed spectra of SN Ia

In Sect. 4.2 we noted that most of the synthetic spectra calculated up to now for SN Ia at early epochs show excess flux in the red and infrared wavelength regions compared to observations (cf. Pauldrach et al. 1996, Nugent et al. 1997, Sauer et al. 2006, and Stehle et al. 2005) simply because the lower boundary of the models has not been placed deep enough into the photosphere. As a result the characteristic observed shape of early-time SNe Ia spectra with the steep slope in the continuum between 6000 Å and 8000 Å (cf. Fig. 12) is not reproduced by the synthetic spectra. As a first step in remedying this problem we have run a model, which we will now discuss, where we calculated much deeper into the photosphere.

**The standard accelerated lambda iteration procedure for single lines (ALI).** By replacing in Eq. 12 the total line source function $S_L(r)$ with $S_m(r)$, the source function of a single line, the ansatz of the standard accelerated lambda iteration (ALI) procedure is recovered. Although the corresponding procedure is not designed to solve the problem of the mutual interaction of strong spectral lines of different ions, it presents at least a robust solution for each single line (cf. Pauldrach & Herrero 1988), and, as is shown on the left hand side of Fig. 13, our first comparison to the observation presented by this model reproduces already the observed steep slope in the continuum in between 6000Å and 8000Å.

However, apart form this success the calculated synthetic spectrum resulting from this single-line ALI model does not at all represent in total the characteristics of a realistic model atmosphere. This is not simply because the density structure of the W7 model, which lies at the basis of this synthetic spectrum, is inadequate: we will show later that a more sophisticated NLTE model yields much more realistic synthetic spectra with this explosion model. The primary reason for this drawback is connected to the flux conservation of this model which is at a miserable 50%. From comparison of different models we realized that the amount of lost flux is proportional to the difference in the spectral energy distribution between the inner boundary and the emergent flux. This behavior reflects the fact that the standard iteration scheme that connects the mutual interaction of strong spectral lines with the "pseudo-continuum", the level populations, and the radiation field, produces exactly the problem we have discussed in Sect. 4.2; and as noted there, the stan-dard ALI procedure for single lines does not automatically solve this problem as was previously thought.

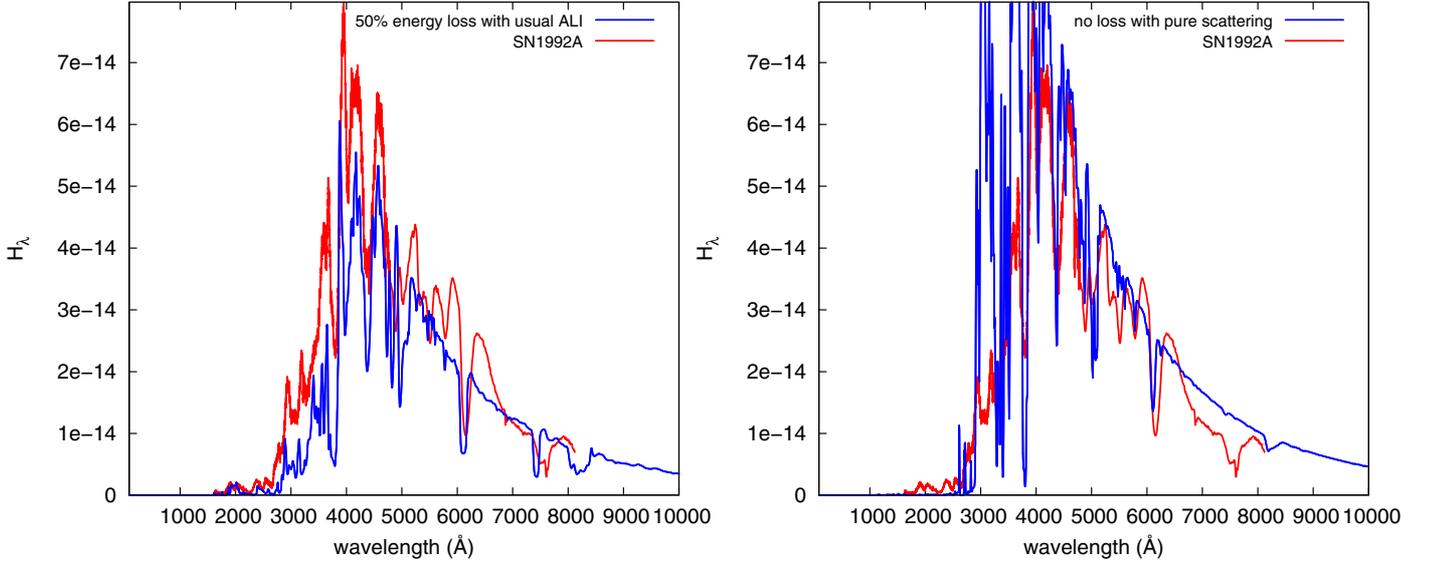

**Fig. 13.** Left: Model spectrum computed using the standard single-line ALI approach for the radiative rates. The iteration is stable but does not converge. Instead, due to the mutual interaction of strong spectral lines of different ions, the model becomes stuck in a state where approximately 50% of the energy disappears. Right: Spectrum from a model where all line opacities and emissivities are treated as purely coherent scattering. Here the flux is conserved, but too much energy is emitted in the blue part, in contradiction to the observations.

A scattering approach for the description of the mutual interaction of strong lines. A straightforward way to overcome the stagnating iteration scheme is to treat strong opacities as scattering opacities, since an intrinsically energy-conserving, selfconsistent solution of scattering (in which the emission is simply $\eta_\nu = \chi_\nu J_\nu$) is possible within the radiative transfer itself, without feedback from the rate equation system.

A comparison of the synthetic spectrum from such a model to the observed spectrum of SN 1992A is shown in the right-hand panel of Fig. 13. Although the scattering procedure has now led to a model that is indeed flux conserving, the spectrum does not fit the observations, being much too blue. Since only true opacities[22] can transfer energy from the blue part to the red part of the spectrum this example clearly shows that the ansatz to solve the radiation transport with scattering does work for the flux conservation but is in no way a physically adequate solution to the problem.

It is noteworthy that although our present calculations use atomic models that have been significantly improved compared to our earlier calculations from Pauldrach et al. (1996) and Sauer et al. (2006) the appearance of the synthetic spectra in regard to their compatibility with the observations is now markedly inferior to those earlier calculations. This apparent setback can only be caused either by the consideration of energy deposition within the simulated layers of the ejecta, or by the fact that the inner boundary of our current models now lies much deeper than in the earlier models. The latter change was necessary to improve the realism of the description in the red part of the spectrum, but its consequence is that the spectral energy distribution at the inner boundary now differs much more from the emergent spectrum, having its maximum at considerably shorter wavelengths than before. Thus, the correctness of the description of the material in the intervening layers in its ability to transform the radiative energy distribution from one with a substantial UV component at the inner edge to the significantly redder distribution of the emergent flux becomes much more critical. We will investigate both effects in the following sections.

---

[22] "True opacity" is a common term in the radiative-transfer field to denote any opacity that is not a scattering opacity (i.e., where $\eta_\nu = \chi_\nu \cdot J_\nu$, such as Thomson opacity), and which thus can couple (via other atomic processes) to the thermal pool.

### 5.2. Improved results obtained by developing and applying an ALImI-procedure

As the scattering approach does not constitute a physically adequate solution to the problem of the mutual interaction of strong spectral lines, a more realistic description of this effect obviously needs to be developed. However, an exact generalization of the established lambda iteration acceleration technique for individual lines to make it usable for supernova ejecta currently appears out of the question, since this would require a rewrite of the rate equations on a line-by-line basis taking into account all other lines that overlap a particular line at almost all radius points at almost all frequencies in the UV, and would thus require that the rate equations *for all ions* to be solved simultaneously. (Similarly, a complete-linearization approach would need to consider all ions of all elements simultaneously.) This is beyond the numerical capabilities of our current computers. However, an exact generalization of the ALI technique is not actually needed to overcome the problem. All that is required is an adequate correction term which accounts for the mutual interaction of the strong spectral lines of different elements and which unfreezes the "stuck" state of the iteration with the constraint that this corrective term goes to zero when true convergence is achieved.

To establish such a lambda iteration acceleration technique accounting for the mutual interaction of strong lines ("ALImI") we start by rewriting Eq. 12 in the form

$$\overline{J}_m(r) = \overline{J}_m^0(r) - \delta S_{\mathrm{L}}^{(i)}(r), \tag{13}$$

where

$$\overline{J}_m^0(r) = \Lambda(S_{\mathrm{L}}(r, \mu), S_{\mathrm{cont}}(r)), \tag{14}$$

is the classical formal solution of the radiative transfer in the comoving or the observer's frame, and

$$\delta S_L^{(i)}(r) = \Lambda^* \left( S_L^{(i-1)}(r) - S_L^{(i)}(r) \right) \qquad (15)$$

is a correction term that includes the approximate operator $\Lambda^*$ which acts on the inconsistency of the total line source function with respect to the current iteration ($i$) and the previous iteration ($i-1$). Since the properties of this operator determine the stability and convergence behavior of the ALImI iteration process, the choice of this operator is of importance. Based on our experience we opt for a local operator, since only this kind of operator offers a good chance to drive multi-level non-LTE calculations to convergence. We thus approximate $\delta S_L^{(i)}(r)$ by

$$\delta S_L^{(i)}(r) = C_{\nu_l}(\tau_{\nu_l}(r)) \left( S_L^{(i-2)}(r) - S_L^{(i-1)}(r) \right), \qquad (16)$$

a suitable form of $C_{\nu_l}(\tau_{\nu_l}(r))$ being

$$C_{\nu_l}(\tau_{\nu_l}(r)) = 1 - e^{-\tau_{\nu_l}(r)/(\gamma \tau_{\nu_l}^0(r))} \qquad (17)$$

where $\tau_{\nu_l}(r)$ is the local optical thickness in the current iteration scaled by $\tau_{\nu_l}^0(r)$, the optical thickness obtained in the first iteration assuming radiative balance – see below –, and $\gamma$ is a controlling parameter (cf. Pauldrach & Herrero 1988); we have also replaced[23] the inconsistency between the source functions of the current and previous iterations, $S_L^{(i-1)}(r) - S_L^{(i)}(r)$, by that of the two previous iterations, $S_L^{(i-2)}(r) - S_L^{(i-1)}(r)$. It is important to realize that the chosen form of $C_{\nu_l}(\tau_{\nu_l}(r))$ has to ensure that the most problematic optically thick cases, where $\tau_{\nu_l}(r) \gg 1$, are treated in a proper way. Equation 16 verifies this required behavior in the way that the operator replacement $C_{\nu_l}(\tau_{\nu_l}(r))$ becomes 1 in these cases, and the iteration procedure is therefore primarily driven by the full inconsistency between the previously calculated source functions obtained by formal solution of the radiative transfer.

While a solution of the regular radiative transfer Eq. 14 and the application of its result to the radiative rate coefficients Eq. 5 represents the usual $\Lambda$ iteration, the application of Eq. 16 to the radiative rate coefficients gives additional "acceleration" terms[24]

$$R_{ji}^+(r) = 2\pi \left( \frac{n_i}{n_j} \right)^* \int\limits_0^\infty \int\limits_{-1}^{+1} \frac{\alpha_m(\nu, \mu, r)}{h\nu} \frac{2h\nu^3}{c^2} \frac{\delta S_L^{(i)}(r)}{S_m(r)} e^{-h\nu/kT} \, d\mu \, d\nu \qquad (18)$$

---

to be taken into account in the solution of the rate equations[25]

$$n_i \sum_{j \neq i} (C_{ij} + R_{ij}(I_\nu)) + n_i(C_{ik} + R_{ik}(I_\nu))$$
$$= \sum_{j \neq i} n_j(C_{ji} + R_{ji}(I_\nu) + R_{ji}^+(r)) + n_\kappa(C_{\kappa i} + R_{\kappa i}(I_\nu) + R_{\kappa i}^*(r)) \qquad (19)$$

and thus the consistent determination of the occupation numbers affected by the lines that have to be considered (this procedure represents the basis of the ALImI iteration process). Although the resulting solution scheme, which is primarily based on the stepwise decreasing influence of the acceleration terms $R_{ji}^+(r)$, turned out to be stable and to converge quickly when applied to the spectral line formation calculations of SNe Ia envelopes, the choice of proper initial conditions is still of importance for the fast convergence of the iteration cycle, and they are decisive for providing an insight into the behavior of the mutual interaction of strong spectral lines of different ions.

The required initial conditions for the iteration cycle concern first of all the starting values for the total line source functions $S_L(r)$ given in Eq. 10. In order to get these starting values (which have to be evaluated at all required frequencies $\nu_l$) we first have to investigate the behavior of the specific line source functions $S_m(r)$. This is done by writing the specific line source functions in the form

$$S_m(r) = \frac{2h\nu^3}{c^2} \frac{1}{b_m(r) \, e^{h\nu/kT} - 1}, \qquad (20)$$

where the $b_m(r)$ terms

$$b_m(r) = \frac{n_i/n_j}{\left( n_i/n_j \right)^*} \qquad (21)$$

describe the departure from LTE. From Eq. 20 it is easily seen that $S_m(r) = B_{\nu_l}(T(r))$ if $b_m(r) = 1$. This however is exactly the case if the radiative rates of the corresponding line are in detailed balance, since in this case collisions are the dominant terms responsible for establishing the populations of the involved levels, and collisions naturally produce LTE ratios $n_i/n_j = (n_i/n_j)^*$ for those occupation numbers[26] even at the intermediate densities (cf. Pauldrach 1987) typical for the supernova ejecta. That the radiative rates of the strongest lines have to be in detailed balance in the deeper part of the envelope is obvious, however, since most of these lines have large optical thicknesses (cf. Fig. 10), and it is furthermore the cumulative effect of these optical line thicknesses that is responsible for the non-convergence of the lambda iteration. Thus, the fact that a great part of the strongest lines is in detailed balance over a considerable part of the envelope produces on the one side the problem of the non-converging

---

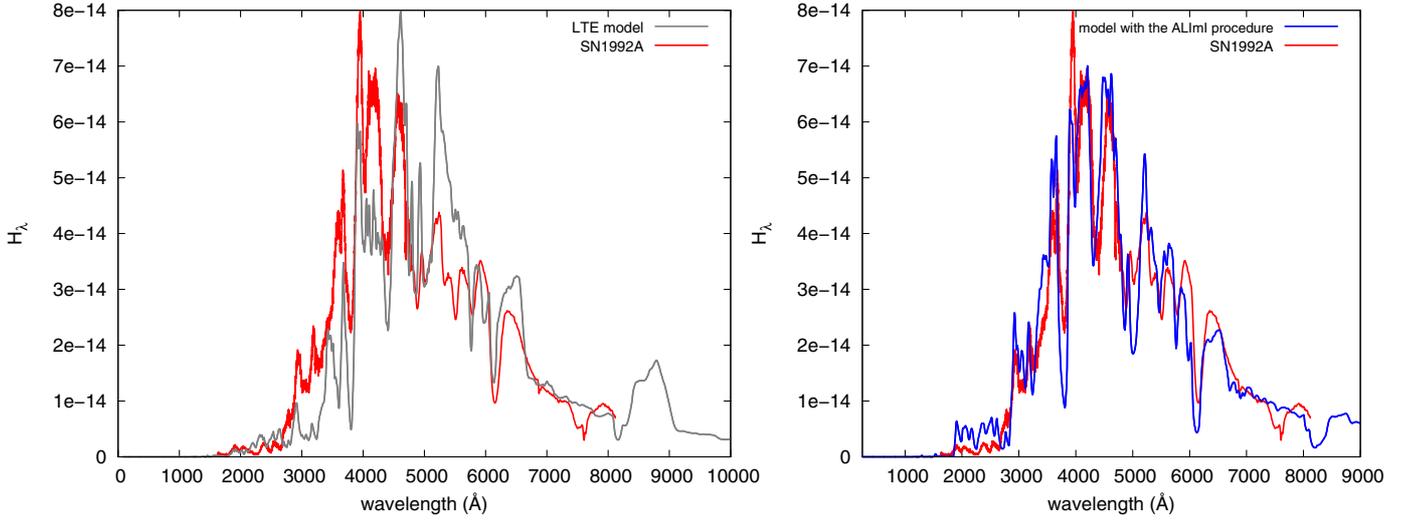

**Fig. 14.** Synthetic spectrum of a consistent ALImI-model (right hand side) and a LTE-model (left hand side) used as a starting configuration for the primary iteration cycle. As it has been the case for the pure scattering model, both models, which are compared to the spectrum of SN 1992A at early epochs (red line, see also Fig. 12), preserve the total luminosity induced at the inner boundary. The shape of the spectrum resulting from the consistent ALImI-model reveals however a significant improvement when compared to the pure scattering model (Fig. 13) and the observations (especially in the blue part of the spectrum – shortward of 5000 Å – the differences to the observation are strongly diminished in the case of the ALImI-model). An improvement of the calculated spectral characteristics is also recognized for the LTE-model revealing that $B_{\nu_i}(T(r))$ presents good starting values for the total line source functions $S_L(r)$. The strong discrepancies of individual line shapes – especially in between 3000–6000 Å – shows however that a simulation technique for diagnostic issues should not be based on LTE-models.

lambda iteration, but on the other side it offers the opportunity to calculate the occupation numbers independently of these lines in a first step. The circumstance that the total optical depth and not the optical depth of each single line is responsible for the extremely small mean free paths of the photons at the line-crowded frequencies magnifies this behavior, because the mean free paths are the quantities which are physically responsible for the behavior that the involved lines represent a detailed balancing of their rates. This means that even lines which would not represent such a behavior by means of their own influence are forced to this state by means of the influence of all the other lines connected to the considered frequency.

As a result of these special circumstances $B_{\nu_i}(T(r))$ represents not only good starting values for the individual line source functions, but also for the total line source functions

$$S_L(r) \lesssim B_{\nu_i}(T(r)). \tag{22}$$

That this behavior is indeed close to the real case is shown in the left panel of Fig. 14, where the synthetic spectrum of the LTE model we use as start model in the new iteration scheme is also compared to observations. Although the shape of the spectrum turns out to be an improvement when compared to the scattering model and the observations, the severe discrepancies found for several spectral lines reveal that this model certainly does not represent a final model to be used for diagnostic purposes. However, the overall shape of the spectrum shows that $B_{\nu_i}(T(r))$ represents an excellent starting approximation for the iteration.

Our procedure utilizes the results of this investigation by using $S_L(r) = B_{\nu_i}(T(r))$ as start values for the first iteration. With a value of $\gamma = 10^6$ we then perform $\Lambda$-iterations for the next two iterations to obtain start values for the $\delta S_L^i(r)$ terms (cf. Eqs. 13–18). Using the additional "acceleration" rates (Eq. 18) we then solve the complete accelerated $\Lambda$-iteration cycle for mutual interaction by systematically reducing the value of $\gamma$ down to $10^{-4}$ until the model converges.

The synthetic spectrum obtained as a result of applying this ALImI procedure in the NLTE model show a striking improvement in reproducing the features of the observed SN Ia spectra (see the right-hand panel of Fig. 14), not only in regard to the overall shape of the spectrum but also in the details of many of the characteristic strong lines of low-ionized intermediate-mass elements. The differences in the emergent spectrum in the UV between this model and the one using the standard single-line ALI (left-hand panel of Fig. 13) clearly shows that the deficiencies of the old model are indeed due to a hidden form of a lambda iteration deadlock caused by the mutual interaction of strong spectral lines inherently connected to line blocking.

For the computation of synthetic SN Ia spectra this effect is of great importance, because lines not only shape the spectral flux distribution directly via blocking, but also indirectly through the resulting influence on the ionization structure.[27] The drastic effects of line blocking on the ionization and excitation and thus the emergent flux can be verified by comparing the calculated incident flux obtained at deeper layers with the emergent flux (see Fig. 15). The huge change in the UV from the inner boundary to the outer layers is due to the increasing contribution of the lines to the opacity in the outward direction. It is an essential by-product of this result that the observed line features in the optical are by far not reproduced in the inner part, although the local temperatures there are not very different from the values in the outer regions. The two strongest spectral features in the inner part are due to Fe III lines, indicating that the ionization is much higher there, just because of the excessive UV flux which is still present in this deeper layers. This behavior nicely illustrates what we have stated before, namely that the ionization balance depends almost entirely on the strength of the ionizing UV flux and that this influence can be traced by the spectral lines in the observable part of the spectrum.

---

[27] As is shown in Fig. 15 line blocking is most effective in the wavelength range of the peak intensity of the radiation emitted from deeper layers of the photosphere – since the photospheric temperature is close to 20 000 K at those layers (cf. Fig. 19), and since the ionization range is still not high in this temperature regime, line blocking is due mostly to species such as Fe II–IV, Co II–IV, and Si II–IV, which have an enormous number of spectral lines in the near-UV (1000–3500 Å); this means that the near-UV is the spectral part where the line blocking process dominates, as can be verified from Fig. 15.

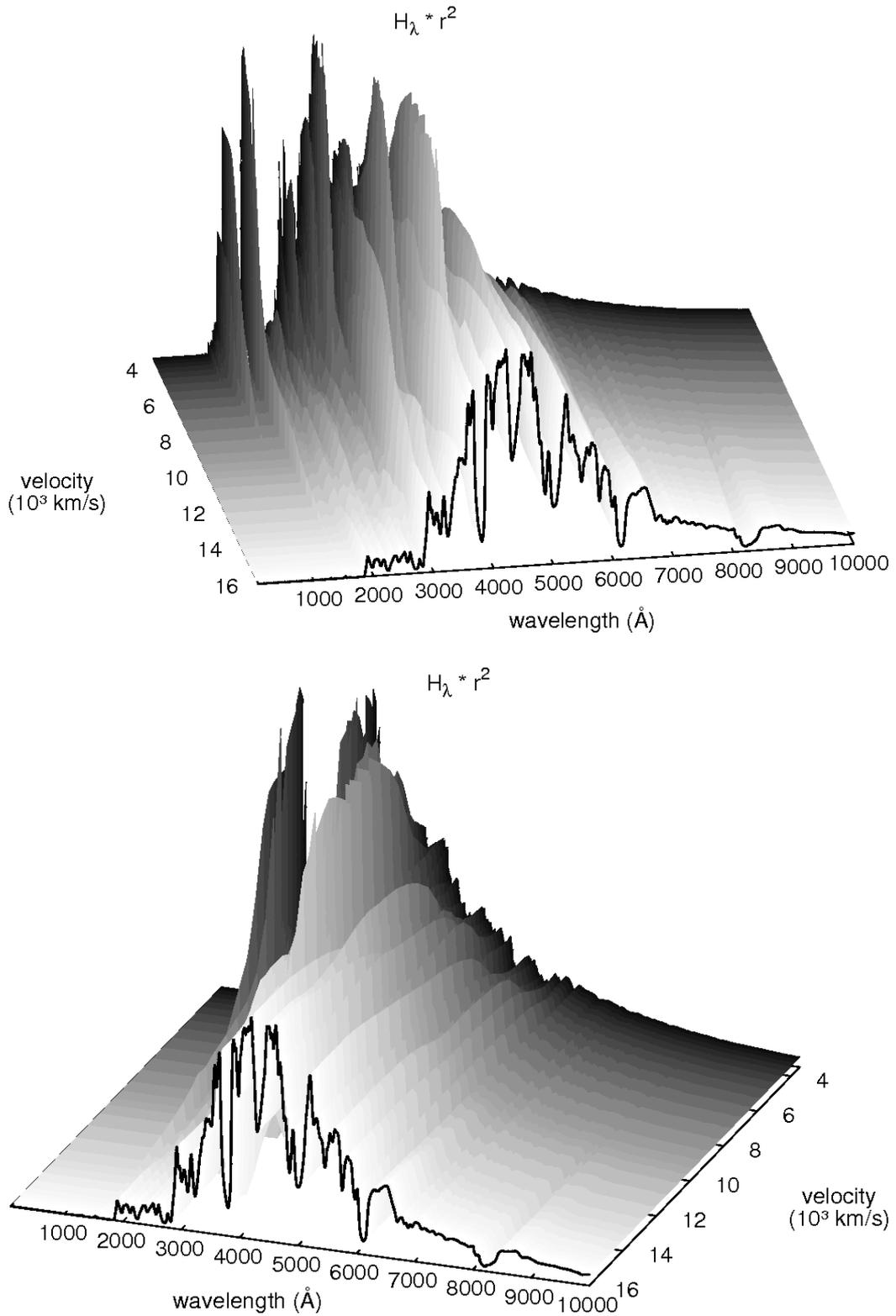

**Fig. 15.** Two different views of the Eddington flux $H_\lambda$ times the square of the radius $r^2$ as a function of wavelength and velocity, displayed from the inner boundary to a radius equivalent of 16000 km/s, for a model with the consistent ALImI-method. Despite the dramatic change in the spectral distribution of the flux, the luminosity (the frequency-integrated flux times the surface, $\int r^2 H_\lambda \, d\lambda$) is constant as a function of radius to within a few percent. The flux in the UV between 1000 Å and 3000 Å is blocked due to the large optical depth of the pseudo-continuum (shown in the upper panel). This radiation is absorbed and reemitted in the optical wavelength range from 3000 Å to 6000 Å (lower panel) where the opacity is smaller and the radiation can escape the atmosphere.

Thus, one of the biggest challenges we have found in the modeling of the radiative transfer in SN Ia is the fact that the radiative energy in the UV has to be transferred into the optical regime in order to be able to leave the ejecta. In calculating deeper into the ejecta, the characteristic shape of the spectral distribution of the radiation at the inner boundary has a maximum intensity in the blue compared to the escaping radiation seen in the observed spectrum. This problem becomes progressively more severe the deeper the inner boundary of the model lies. In "normal" stellar atmospheres the spectral shift between inner and outer boundaries principally exists as well, but stellar objects have a dominating continuum connected to the temperature where one and the same physical process can absorb in the blue and emit in the red, thus preventing such a problem from appearing. Using the standard single-line ALI procedure the conversion of radiation through strong lines does not happen. (As we will see in the next section, the consideration of energy deposition within the ejecta does not change the overall behavior of the system.) As we have shown, however, our ALImI-procedure solves this principal problem and thus constitutes a foundation for more refined models, such as those including energy deposition.

To summarize this section:

- The problem we have had with a deadlocked lambda iteration is the result of us having previously employed the usual line-by-line ALI (but the standard complete-linearization scheme shows from its structure the same weakness).
- We describe a multi-line ALI applicable in the observer's frame method we use (but the procedure can also be applied in the comoving frame).
- The problem of overlapping lines is a general problem and must be solved by all codes.

### 5.3. Current best results obtained by applying time-dependent energy deposition rates

To simulate more realistic spectra for the photospheric epoch requires the construction of consistent models which not only link the results of the nucleosynthesis and hydrodynamics with the computation of synthetic spectra obtained via the assumption that the total luminosity originates from below the boundary of the simulation volume, but which also make use of the results of light curve calculations. This concerns the radioactive energy rates of the decay products of $^{56}$Ni and $^{56}$Co, which are deposited in the ejecta and from which the total luminosity of SNe Ia results (cf. Sect. 3). Thus the luminosity of the model must be specified only partly at the inner boundary to account for the radiative energy deposited below the simulation volume of the radiative transfer, and the energy deposition within the ejecta must be treated explicitly. If this physical behavior is described correctly, it could of course have important implications for the results obtained from the steady-state non-LTE simulations of the envelope.

For a first comparison to observations based on such a theoretically complete description we have again chosen the photospheric epoch of SN Ia, because the spectra of this epoch (lasting for about a month after explosion) contain the most useful information on the energetics connected to the explosion. Our primary objective in this regard is to analyze differences in the characteristic shapes of the SN Ia spectra obtained with and without energy deposition.

With respect to this objective the time-dependent radiative transfer, which determines the energy deposition rates $E_{dep}(v, t)$, the luminosity $L_{IB}(t)$ to be used at the inner boundary of the non-LTE model (for this part of the luminosity we use the inner boundary condition described by Sauer et al. 2006), and the synthetic light curve, is solved in a first step (cf. Sect. 3). As it has already been verified via a successful comparison of our calculated synthetic light curve resulting from the time-dependent radiative transport to an observed light curve (cf. Fig. 7) that the total energy deposition rates are described accurately by this procedure, *improved snapshots of synthetic SN Ia spectra resulting from the non-LTE simulations can be calculated in a second step by treating the calculated rates of the energy deposited in the outer part of the ejecta as emissivities $\eta_v^{dep}(v, t)$ in the radiative transfer and in the energy equation* (cf. Fig. 1). (The energy-deposition simulations track the transport of gamma photons and positrons, but after deposition that energy must eventually escape as optical and UV photons, with a yet to be determined spectral distribution.)

This distribution can in principle be computed by explicitly modeling the microphysics in the MeV and keV regime down to the eV regime in which the UV and optical synthetic spectra are calculated. But since the $\gamma$-photons generate fast electrons through Compton scattering which transfer the energy back to the radiation field either through Bremsstrahlung or ionizing processes (Sutherland & Wheeler 1984; Swartz et al. 1995), the electron distribution quickly thermalizes and the portion of radiation which is produced during this interaction and which represents exactly the original deposited energy also almost immediately goes into equilibrium with the already existing radiation field (as long as the pseudo-continuum does not become optically thin in the complete frequency range – cf. Fig. 10). Thus, the deposited energy appears in the form of optical and UV photons whose spectral distribution is determined and shaped by a myriad of non-LTE interactions in a similar manner as the radiation that has already been deposited and processed by the same type of interactions in the layers below. The normalized spectral distribution of the deposited emissivity is therefore well described by the flux enhancement $\Delta H_v$ resulting from the calculated energy deposition rate. Although it would be desirable to have computations which explicitly model the microphysics in the MeV and keV regime in the same code as that used to model the eV regime, they are not urgently required for those epochs where the electrons quickly thermalize and the radiation produced by these microphysical interactions also goes into equilibrium with the already existing radiation field.

The spectral energy distribution that arises is thus a result of the many interacting microphysical processes which together describe the flow of energy through the ejecta. The approximation that the quantum of energy that is deposited in a given layer will be shaped in a similar manner as the radiation that has already been deposited and processed by the same type of interactions in the layers below is therefore a much better approach than, for instance, naively assuming a Planck spectral distribution (which would imply detailed equilibrium). Moreover, in supernovae Ia envelopes the optical depth is strongly frequency-dependent, and in outer layers where some frequencies may already be optically thin (the flux comes from deeper layers in these regions), others might still be optically thick and through the matter/radiation field interactions at these frequencies still process deposited energy. A direct consequence of this behavior are the low infrared

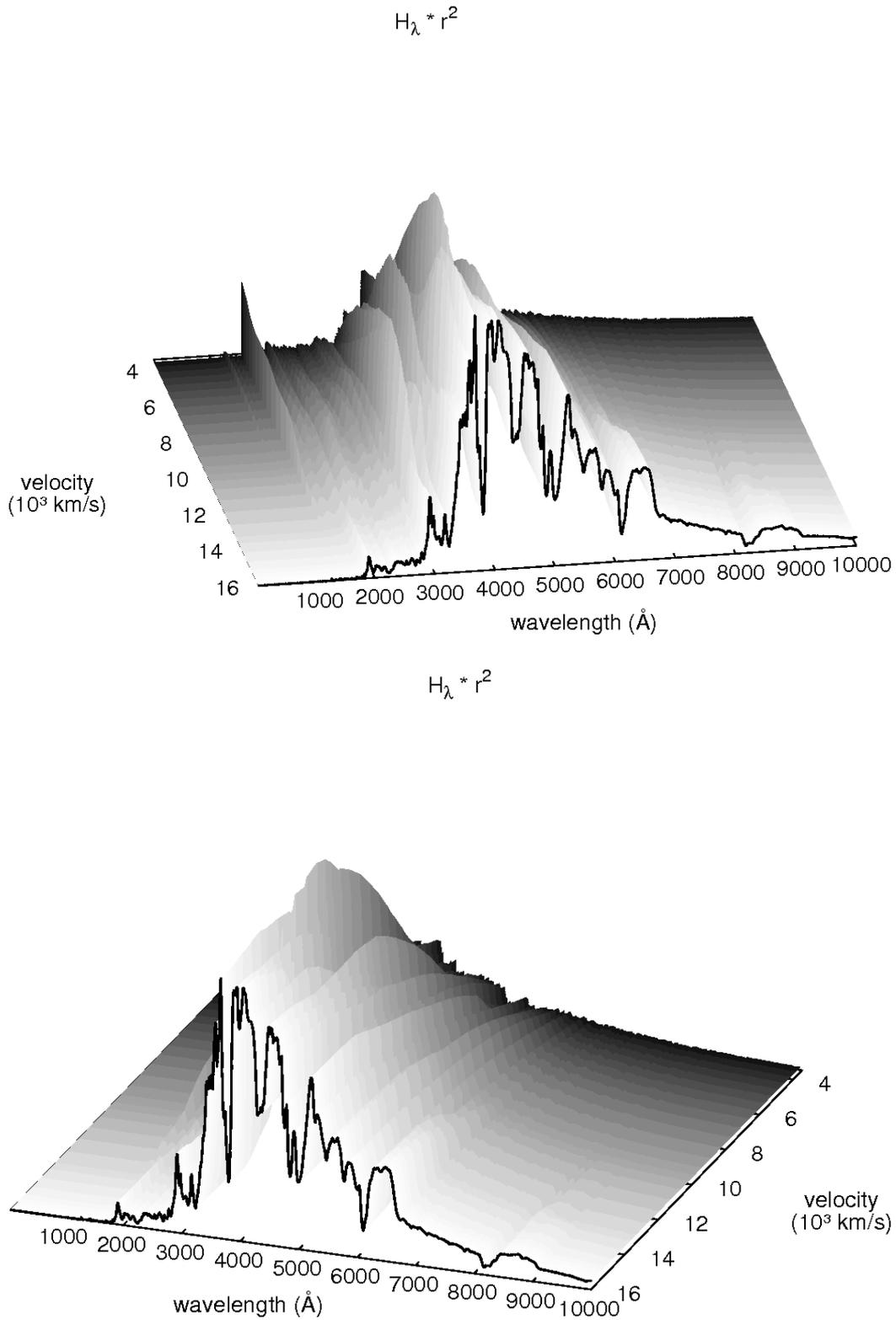

**Fig. 16.** Two different views of the Eddington flux $H_\lambda$ times the square of the radius $r^2$ as a function of wavelength and velocity, displayed from the inner boundary to a radius equivalent of 16000 km/s, for a model with energy deposition. Compared to the energy-conserving atmosphere shown in Fig. 15 the change in the spectral distribution of the flux does not appear quite as dramatic, because the flux produced as a result of the energy deposited in the atmosphere from the radioactive decay is added gradually with the local spectral characteristics. Thus, the blocking effect in the UV is less prominent (upper panel), but the emission in the optical is comparable to the model without energy deposition (lower panel).

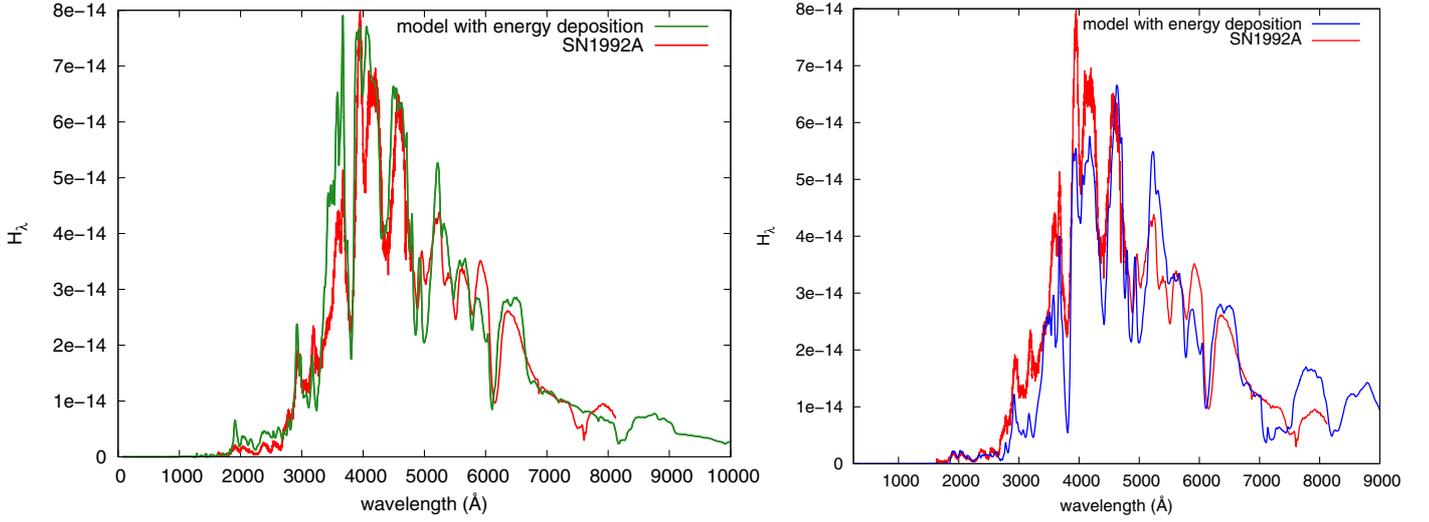

**Fig. 17.** Synthetic spectra for consistent ALImI-models calculated with energy deposition compared to the spectrum of SN 1992A (red line). The spectrum on the right hand side belongs to a model where the energy deposited in the ionization rates has in comparison to the model whose spectrum is shown on the left hand side been reduced by 50% in order to show the influence of this energy component on the observed spectral features. As is recognized, the difference to the observation in the blue part of the spectrum (shortward of 3500 Å) is strongly diminished in this case.

flux levels (compared to a black body) which are indeed observed in SN Ia spectra.

The behavior of the flux enhancement $\Delta H_\nu$ resulting from the deposited energy can be discovered from the energy equation by considering that the energy deposited in a thin radius shell does not influence the temperature of this shell at such a rate that the condition of radiative equilibrium (first integral on the right hand side of the energy equation displayed in Fig. 1) does not remain approximately fulfilled for this shell. One thus gets from the energy equation and Eq. 1 the relation

$$\Delta\left(r^2 H_\nu(v,t)\right) = \eta_\nu^{\mathrm{dep}}(v,t)\, r^2 \Delta r = \frac{1}{4\pi}\, f_\nu(v,t)\, E_{\mathrm{dep}}(v,t)\, r^2 \Delta r \quad (23)$$

which, when integrated over frequency, serves as the foundation of the modified flux correction procedure discussed in Sect. 4.

The frequency integral of Eq. 23 thus specifies the normalization of the emissivities that result from energy deposition (and states that in a stationary model all deposited energy $E_{\mathrm{dep}}$ must lastly appear as an increment $\Delta H$ in the total flux). The spectral distributions $f_\nu(v,t)$ of the emissivities, however, are determined iteratively by the radiative transfer which treats the additional emissivity $\eta_\nu^{\mathrm{dep}}$ together with all other emissivities and opacities and the flux transported out from the lower shells to yield the flux enhancement $\Delta H_\nu$ by the considered shell. The normalized distribution functions of the deposited energy packets can therefore be approximately described by

$$f_\nu(v,t) = \frac{\Delta(r^2 \bar{H}_\nu(v,t))}{\Delta(r^2 \bar{H}(v,t))}, \quad (24)$$

where the flux distribution $\bar{H}_\nu = \int H_{\nu'} \cdot g(\nu - \nu')\, \mathrm{d}\nu'$ is smoothed with a Gaussian kernel $g(\nu)$ using a width of several 10 Å to prevent unrealistic magnification of the flux at individual frequency points.

To summarize our approach:

- The process of gamma-ray deposition which leads to fast electrons, which in turn leads to non-thermal ionization, which in turn leads to radiation is treated in the frame of our procedure by using the method of Cappellaro et al. (1997).

- This procedure gives us locally additional radiation packages, which at no point are assumed to thermalize, but which are added to the radiation field in the form of emissivities (and also, in accord with the rules for treating moments of the Boltzmann equation, to the energy equation).

- This non-thermalized radiation now affects all interactions which are part of the non-LTE model and is shaped by all of these non-LTE interactions. The spectral energy distribution of this extra radiation is not specified a priori. But since the amount added in each radius shell is small, it is assumed not to dominate the local radiation field, but instead to quickly go into equilibrium with the existing radiation field that is already the result of many interacting microphysical processes which together describe the flow of energy through the ejecta.

Thus, there are two approximations involved in our procedure:

- The microphysics in the MeV and keV regime is not explicitly modeled in the same code as that used to model the eV regime. [But these computations are not urgently required for those epochs where the electrons quickly thermalize and the radiation produced by these microphysical interactions also goes into equilibrium with the already existing radiation field which determines the non-LTE populations.]

- The spectral distribution of the energy-deposition, which can in principle be tracked by the transport of gamma photons and positrons, has yet to be determined. [But since the portion of radiation which is produced during this interaction almost immediately goes into equilibrium with the already existing radiation field, the spectral distribution of the additional radiation packages (for those epochs where the pseudo-continuum does not become optically thin in the entire frequency range) can be related to the required flux enhancement corresponding to the deposited energy.]

The resulting evolution of the spectrum as a function of wavelength and radius on its run through the envelope – displayed from the inner boundary to a radius equivalent of 16 000 km/s – for a model with this simulation of the energy deposition is shown in Fig. 16. Although the luminosity emitted below the inner boundary of this model is still significant compared to the flux originating from the γ-ray energy deposition above that boundary (here we also account for the photons that have been generated at earlier epochs, but have been trapped by the large opacities), the radiative energy generated within the expanding medium is obviously not negligible, since it modifies the process of radiative transfer significantly. As is verified by a comparison to the flux-conserving atmosphere shown in Fig. 15 (in which the total luminosity is injected at the inner boundary and no deposition is considered above this layer) the change in the spectral distribution of the flux with depth does not appear quite as dramatic, because the flux produced as a result of the energy deposited in the atmosphere from the radioactive decay is added gradually with the local spectral characteristics. The blocking effect in the UV is therefore especially at larger depths (from 4 000 to 10 000 km/s) less prominent, although the emission in the optical is comparable to the model without energy deposition.

Compared to the outermost regions the radiation field in the innermost part nonetheless has a significant UV component, and is not at all comparable to that of a thermal (blackbody) emitter. The latter fact is also noticeable in those wavelength regions which are not significantly affected by line opacities (namely, the red and infrared wavelength regimes, cf. Fig. 10). As opacities are required in the outer parts of the ejecta for down-scattering the products of the γ-photons further, energy deposition can not generate emission in these wavelength regions and this missing emission (compared to a black body) results in the characteristic shape of SN Ia spectra where the slope of the emergent spectrum is generally steeper towards the red and infrared wavelength regimes than the slope of a corresponding black body. This result clearly indicates that *the radiation resulting from the deposition of the γ packets is far from being thermalized* (cf. Hoffmann et al. 2013).

Not only are the characteristic overall shapes of SN Ia spectra at early times reproduced but also the prominent line features (cf. Fig. 17). As the products of the γ-deposition transfer their energy to the radiation field through a mixture of Bremsstrahlung and ionizing processes, it is interesting to see how a change of the relative ratios of these processes influences the emergent spectrum. For this purpose we show in Fig. 17 a comparison of two spectra, where the right-hand spectrum was obtained from a model in which the influence of the additional emissivities on the ionization rates has been reduced by 50% compared to the left-hand standard model. (Since we do not model the microphysics of interactions of ions and fast electrons/gamma photons in detail, we have a priori no detailed knowledge of how large this nonthermal effect on the ionization rates actually is. We have therefore experimentally modified the normalized distribution functions to vary their influence on the ionization rates in order to investigate possible variations on the models.) Apart from minor differences in some spectral features, a significant characteristic of this comparison concerns the difference to the observed spectrum blueward of 3 500 Å, which is strongly diminished in the model where the influence of the energy deposited in the ionization rates has been reduced. This result indicates strongly that an appropriate treatment of the effects of the energy deposition is not dispensable for the computation of the radiative transfer of SN Ia envelopes, and it indicates that the part of the luminosity which is deposited within the expanding medium and which depends on the physical state of the radially distributed material in the ejecta is already described on a quantitative level.

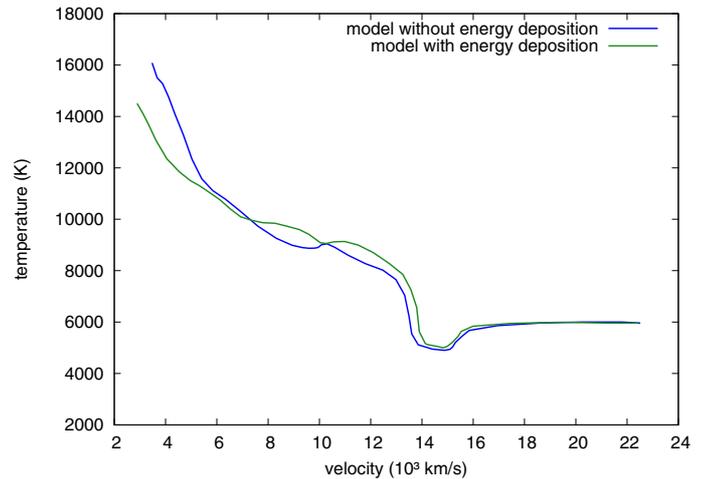

**Fig. 19.** Temperature versus depth for a model with and a model without energy deposition. Because the total flux must already be present at the inner boundary of the latter model, a higher temperature gradient in the inner region is obtained, showing that not all heating processes are incorporated physically correctly in this case. The fact that except from this innermost region the gradients of the consistently calculated temperature structures are not much different for both types of models is the reason that the characteristic shapes of the SN Ia spectra obtained with and without energy deposition are quite similar (cf. Fig. 18).

For the computation of the emergent synthetic SN Ia spectra at early times the treatment of the energy deposition is, however, not entirely critical, as can be verified from Fig. 18 which shows our synthetic spectra for consistent ALImI-models calculated with and without energy deposition compared to the observed spectrum of SN 1992A. The circumstance that the characteristic features and the overall shape of the SN Ia spectra obtained with and without energy deposition are quite similar is on the one hand based on the fact that a great part of the deposited energy is at these times indeed already injected at the inner boundary of the atmosphere, and on the other hand based on the fact that it is primarily the drastic line blocking effect which reduces the intensity of the radiation field in the UV throughout the envelope and thereby shapes the emergent spectrum via its influence on the ionization structure. This finding is also substantiated by the result that the gradients of the consistently calculated temperature structures shown for the models with and without energy deposition in Fig. 19 are not much different, except for the unobservable deepest layers.

Although there is still need for further improvements regarding, for instance, the use of state-of-the-art hydrodynamical explosion models, we conclude from our discussion above that Fig. 18 exemplifies the status quo of calculated synthetic spectra of SNe Ia at early phases, primarily because the model spectra shown reproduce the observed spectrum in a comprehensive way; the overall impression of this comparison is that the method used has reached a level where one may consider the required basic physics to have been treated adequately. With respect to the facilities of spectral analysis we have thus reached the point where radiative transfer models of SNe Ia are based on a similarly serious approach as is generally the case for hot stars with expanding atmospheres, and our present models can therefore be regarded as a basis for the primary goal, namely to tackle the question of whether SNe Ia are "standard candles" in a cosmological sense.

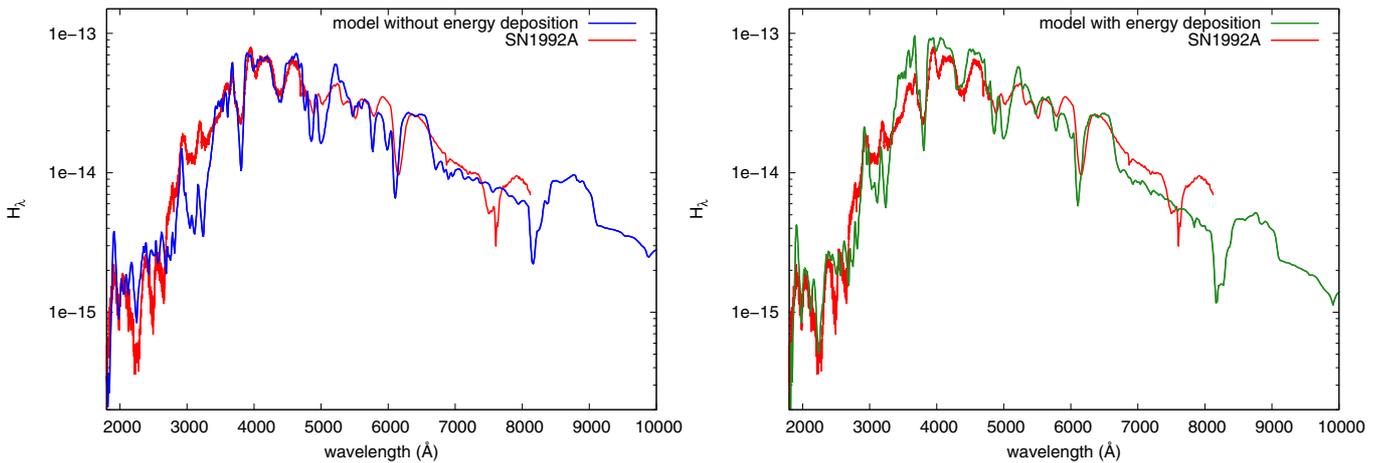

**Fig. 18.** Synthetic spectra for consistent ALImI-models calculated with (right hand side) and without energy deposition (left hand side) compared to the spectrum of SN 1992A (red line). The circumstance that the characteristic features and the overall shape of the SN Ia spectra obtained with and without energy deposition are quite similar is on the one hand based on the fact that a great part of the deposited energy is indeed induced at the inner boundary of the atmosphere, and on the other hand based on the gradients of the consistently calculated temperature structures which – apart from the innermost regions – are not much different for both types of models (cf. Fig. 19).

## 6. Summary and conclusions

Type Ia supernovae, which are the result of a thermonuclear explosion of a compact low mass star, are currently the most widely used probes for cosmological distances. But with respect to diagnostic issues these objects have not yet reached their level of significance in modern astronomy, because neither the explosions nor the radiation emerging from the envelopes of SN Ia are understood in detail. Consequently, serious caveats for the cosmological interpretation of distant supernovae exist. Spectroscopy is the tool that can give answers to the numerous questions which rule this field of work. For example, "Are SN Ia indeed standard candles, or is there some evolution of the supernova luminosity with cosmological age?" is one of the key questions that can only be answered convincingly by searching for spectral differences between local and distant SN Ia. Unfortunately, at present, we do not have a final picture of the exact physical processes which take place in SN Ia explosions and how the radiation released from them has to be treated in detail. In this regard one of the biggest challenges we have found in the modeling of the radiative transfer in SNe Ia is the fact that the radiative energy in the UV has to be transferred practically only via spectral lines into the optical regime in order to be able to leave the ejecta. In "normal" stellar atmospheres the spectral shift between inner and outer boundaries principally exists as well, but stellar objects have a dominating continuum connected to the temperature where the same physical process can absorb in the blue and emit in the red, thus preventing such a problem from appearing.

In this paper we have pointed out and investigated the most necessary steps required for a quantitative analysis of SN Ia spectra based on a serious approach for expanding atmospheres characterized by a sophisticated treatment of the strong spectral lines which interact in an interwoven way with a "pseudo-continuum" that is itself entirely formed by these Doppler-shifted spectral lines, and characterized by the treatment of radioactive energy rates deposited in the ejecta. This approach has been used to obtain a consistent solution of the full non-LTE rate equations along with a detailed solution of the radiative transfer.

For this we had to first uncover a *hidden form of a lambda iteration deadlock*, and we had to demonstrate that a more consistent description of the line processes involved in the simulations of expanding atmospheres of supernovae Ia is required.

Based on these findings we deduced with the ALImI-procedure (accelerated lambda iteration for the mutual interaction of strong spectral lines) a method which solved the problem of the frozen iteration scheme using the conventional single-line ALI treatment, where the iteration is stable but does do not converge, the model becoming stuck in a state where approximately 50% of the flux disappears. The application of our ALImI-procedure showed that with this more consistent description changes in the energy distributions and line spectra are obtained with a much better agreement with the observed spectra when compared to previous models. That the discrepancies in the UV between observed spectra and synthetic spectra from models using the standard single-line ALI approach for the radiative rates are indeed due to this hidden deadlock of the lambda iteration has thus been shown, with important repercussions for quantitative analyses of the metal-dominated type Ia supernova spectra.

The results of this investigation have furthermore lead to an improved understanding of the formation of the observed spectra with respect to their evolution as function of depth in the atmospheres. By showing the diversity of non-observable spectra obtained at different radii of the SN Ia ejecta it becomes clear that the characteristics of the observed spectra would change completely if we could look deeper into the atmosphere. It is astonishing that observed normal SN Ia spectra at early epochs are so similar to each other even though they *could* theoretically vary much more if we could in just some cases just look deeper into the ejecta. This is not altogether self-evident without such depth-dependent spectral modelling and does merit some reflection. Thus this result indicates strongly that the behavior of these objects as a group is very homogeneous with respect to their reaction on the explosion and the structure of their progenitors. By analyzing the observed spectra of SNe Ia we thus get not only information about the emergent energy distribution of these objects, but we can also extract information about numerous unobservable SN Ia spectral characteristics formed in much deeper layers, where the real continuum is still optically thin, from them, since the unobservable spectra are in a complex way the foundation for the observed ones, and are thus automatically analyzed in parallel to the emergent SN Ia spectral features.

To analyze the spectra in the photospheric epoch (lasting for about a month after explosion) which contain useful information on the energetics of the explosion, requires the construction of consistent models which not only link the results of the nucleosynthesis and hydrodynamics with the calculations of synthetic spectra but also with those of the light curves. In this regard the radioactive energy rates of the decay products of $^{56}$Ni and $^{56}$Co, which are deposited in the ejecta and which characterize along with the non-LTE rate equations a consistent solution of the radiative transfer, are especially of importance. These rates have to be calculated via time-dependent consistent solutions of the populations of the atomic levels, the continuum and line transfer, and the treatment of the energy deposition, and they have to be incorporated in the simulations of the synthetic spectra in a consistent manner.

In our present approach the amount and distribution of the time-dependent deposited energy is quantified via the application of a Monte Carlo code which simulates within the ejecta the propagation and absorption of the γ photons and positrons resulting from the radioactive decay chains and which has been tested by a comparison of the resulting synthetic light curves with corresponding observations. The comparison shows clearly that our simulation fits the observed steep increase in brightness during early times, the steep decrease after maximum brightness, and the more shallow decrease up to 300 days after explosion. This result verifies that our procedure describes the total energy deposition rates derived from a time-dependent treatment correctly and that these quantities can therefore be used for the computation of synthetic SN Ia spectra.

On basis of the ALImI-procedure and the energy deposition rates our full non-LTE treatment for the simulation of the SN spectra revealed two important results. First, the radiation generated within the expanding medium by energy deposition is important, since it modifies the mechanism of the energy transfer in an intricate way. An inspection of the formation of the emitted spectrum with regard to its evolution as function of depth in the atmosphere showed that, compared to the outer regions, the radiation field in the inner parts has a much bluer, but non-Planckian, characteristic indicating that the radiation from the deposition of γ rays has not been thermalized. It also showed that no emission from down-scattering of γ-photons can be generated further out in the ejecta in those wavelength regions that do not have significant line opacity, and this effect results – both in the case where energy deposition is considered in the radiative-transfer simulation as well as in the case where the total luminosity is already injected at the inner boundary – in the characteristic shape of SN Ia spectra in the red and infrared wavelengths where the slope of the emergent spectrum is generally steeper than the slope of a corresponding blackbody (cf. Fig. 9 of Hoffmann et al. 2013). The finding that the characteristic shapes of the SN e Ia spectra obtained with and without energy deposition are quite similar is correlated with the gradients of the consistently calculated temperature structures being not much different for both types of models.

These results indicate that the method we describe here has now reached a quantitative level and that the basic physics have been treated adequately. Since we have not fine-tuned our model, the fact that the synthetic spectra reproduce the general features of "normal" type Ia supernovae also indicates that the assumed radially structured element distribution in the ejecta is, together with the velocity and density structure, fairly realistic for a generic SN Ia. The latter result is somewhat surprising, since our simulations here have been based on the relatively old W7 explosion model. Thus there is certainly room for further improvements regarding especially the use of newer, state-of-the-art hydrodynamic explosion models. But with respect to our present results we do not expect to find severe differences on basis of such simulations. We conclude that our radiative transfer models of SN Ia now produce synthetic spectra which are of similar quality as those produced by our models for expanding atmospheres of hot stars, and thus the facilities required for spectral analyses to investigate differences between normal and peculiar SN Ia, or possible differences between distant and nearby supernovae Ia, have now been developed.

*Acknowledgements.* We wish to thank our colleague P. Mazzali for providing us with a basic solver routine for light curves. We also thank an anonymous referee for helpful comments which improved the paper. P. Hultzsch thanks the Max-Planck Gesellschaft for support and hospitality. This work was supported by the Deutsche Forschungsgemeinschaft under grant PA 477/7-1.